 \documentclass[apj]{emulateapj}
\usepackage{color}
\usepackage{epsfig}
\usepackage{amsmath}
\usepackage{longtable}
\usepackage{graphicx}

\DeclareMathOperator\erf{erf}

\shorttitle{Particle Acceleration in Relativistic Magnetic Reconnection}
\shortauthors{Guo et al.}

\doublespace
\begin{document}

\title{Particle Acceleration and Plasma Dynamics during Magnetic 
Reconnection in the Magnetically-dominated Regime}

\author{Fan Guo\altaffilmark{1}, Yi-Hsin Liu\altaffilmark{2}, 
William Daughton\altaffilmark{1}, and Hui Li\altaffilmark{1}}

\altaffiltext{1}{Los Alamos National Laboratory, 
Los Alamos, NM 87545, USA}

\altaffiltext{2}{NASA Goddard Space Flight Center, Greenbelt, MD 20771, USA}


\email{guofan.ustc@gmail.com}

\begin{abstract}
Magnetic reconnection is thought to be the driver for many
explosive phenomena in the universe. The energy release and
particle acceleration during reconnection have been proposed 
as a mechanism for producing high-energy emissions and 
cosmic rays. We carry 
out two- and three-dimensional kinetic simulations to 
investigate relativistic magnetic reconnection and the 
associated particle acceleration.
The simulations focus on electron-positron plasmas starting 
with a magnetically dominated, force-free current sheet 
($\sigma \equiv B^2/(4\pi n_e m_e c^2) \gg 1$). 
For this limit, we demonstrate 
that relativistic reconnection is highly efficient at 
accelerating particles through a first-order Fermi process
accomplished by the curvature drift of particles along the
electric field induced by the relativistic flows. This 
mechanism gives rise to the formation of hard power-law spectra 
$f \propto (\gamma-1)^{-p}$ and approaches $p = 1$ for 
sufficiently large $\sigma$ and system size. Eventually most of the
available magnetic free energy is converted into nonthermal
particle kinetic energy. An analytic model is presented to 
explain the key results and predict a general condition for 
the formation of power-law distributions. The development of 
reconnection in these regimes leads to relativistic inflow 
and outflow speeds and enhanced reconnection rates relative 
to non-relativistic regimes. In the three-dimensional 
simulation, the interplay between secondary kink and tearing
instabilities leads to strong magnetic turbulence, but does not 
significantly change the energy conversion, reconnection rate, 
or particle acceleration. This study suggests that relativistic 
reconnection sites are strong sources of nonthermal particles, 
which may have important implications to a variety of 
high-energy astrophysical problems.
\end{abstract}


\section{Introduction}

Magnetic reconnection is a fundamental plasma process that
rapidly rearranges magnetic topology and converts magnetic 
energy into various forms of plasma kinetic energy, including 
bulk plasma flow, thermal and nonthermal plasma 
distributions \citep{Kulsrud1998,Priest2000}. It is thought
to play an important role during explosive energy release
processes of a wide variety of laboratory, space and 
astrophysical systems including tokamak, planetary 
magnetospheres, solar flares, and high-energy astrophysical
objects. Relativistic magnetic reconnection is often invoked 
to explain high-energy emissions and ultra-high-energy cosmic 
rays from objects such as pulsar wind nebulae 
\citep[PWNe;][]{Kirk2004,Arons2012,Uzdensky2014}, jets 
from active galactic nuclei \citep[AGN;][]{Pino2005,Giannios2009}, 
and gamma-ray bursts 
\citep[GRBs;][]{Thompson1994,Zhang2011,McKinney2012}. 
In those systems, the magnetization parameter 
$\sigma \equiv B^2/(4\pi n_e m_e c^2)$, is often estimated 
to be much larger than unity $\sigma \gg 1$ and the
Alfv$\acute{e}$n speed approaches the speed of light 
$v_A \sim c$. To explain the observed high-energy emissions, 
often an efficient energy conversion mechanism is required 
\citep[e.g.,][]{Zhang2007,Celotti2008,Zhang2011,Zhang2013}. 
Collisionless shocks, which can efficiently convert plasma 
flow energy into thermal and nonthermal energies in 
low-$\sigma$ flows, are inefficient in dissipating
magnetically dominated flows, where most of the energy is stored in
magnetic fields. In these regimes, magnetic reconnection is the primary candidate 
for dissipating and converting magnetic energy into relativistic 
particles and subsequent radiation. Understanding magnetic
reconnection is also important to solve the so-called 
$\sigma$-problem
\citep{Coroniti1990,Lyubarsky2001,Kirk2003,Porth2013}, where 
strong magnetic dissipation may be required to convert the 
magnetically dominated flow ($\sigma \gg 1$) to a matter-dominated 
flow ($\sigma \ll 1$). However, the detailed physics of 
relativistic magnetic reconnection, including the magnetic 
reconnection rate, energy conversion and particle acceleration,
are not well understood.

\citet{Blackman1994} and \citet{Lyutikov2003} have studied 
the properties of relativistic magnetic reconnection using 
the extended Sweet-Parker and Petschek models.
They found that when $\sigma \gg 1$ the outflow 
speed $u_{out}$ approaches 
the speed of light, and the rate of relativistic magnetic 
reconnection and inflow velocity $u_{in}$ may increase compared to the 
nonrelativistic case. 
This is because of the enhanced outflow density arising 
from the 
Lorentz contraction of plasma passing through the diffusion 
region $u_{in} \propto u_{out} \Gamma_{out}/\Gamma_{in}$,
where $\Gamma_{out}$ and $\Gamma_{in}$ are Lorentz factors
of outflows and inflows, respectively. However, later analysis \citep{Lyubarsky2005} showed
that for a pressure-balanced current layer
($B^2/8 \pi \sim nk (T_i + T_e)$), the thermal pressure constrains
the outflow speed to be mildly relativistic and hence the effect
of Lorentz contraction is negligible. Although the rate of
relativistic magnetic reconnection is reported to enhance
in a number of
studies using different numerical models
\citep{Zenitani2009,Bessho2012,Takamoto2013,Sironi2014,Guo2014,Melzani2014a}, its nature is not clear. This issue is recently
revisited by carefully analyzing results from fully kinetic
two-dimensional (2D) simulations \citep{Liu2015}, which shows 
that the plasma density and pressure around the X-line drop significantly as the initial high-pressure region is depleted
during reconnection. This results in a reconnection region 
with $\sigma \gg 1$ and a relativistic
inflow speed $v_{in} \sim c$. The local reconnection rate 
across the diffusion region is well predicted by a simple 
model that includes the Lorentz contraction. However, the
extension of these results to three-dimensional (3D) kinetic 
simulations was not considered.

Plasma energization during magnetic reconnection has been 
extensively discussed in the literature. However, the primary
acceleration mechanism is still unclear.  \citet{Romanova1992} 
analyzed particle motions in a large-scale reconnection region 
and predicted a spectrum $dN/d\gamma = \gamma^{-p}$ with 
$p = 1.5$ for the pair plasma case. \citet{Litvinenko1999} 
found a solution with a spectral index $p = 2$ when particles 
are accelerated in a  direct electric field associated with 
magnetic reconnection. Using a model for the motions of particles 
in a steady magnetic reconnection region, \citet{Larrabee2003} 
have found strong particle acceleration in the reconnection layer 
and obtained a hard energy spectrum with a spectral index of 
about $p = 1$. The first-order Fermi acceleration in converging 
reconnection inflows has been discussed 
\citep{Pino2005,Lazarian2009,Kowal2012}. \citet{Drury2012} 
studied the acceleration in a reconnection layer including 
energy change in both inflows and outflows and show fluid 
compression is crucial for efficient particle acceleration.
Test-particle approach has been applied to 
interpret the strong particle acceleration responsible for
$\gamma$-ray flares from the Crab pulsar \citep{Cerutti2012}. 
The results show that magnetic reconnection may be the site of 
extreme particle acceleration required to explain high energy
emissions from the Crab flares \citep[see also][]{Cerutti2013}. 
Self-consistent kinetic simulations have been widely used to 
study plasma dynamics and particle energization during magnetic
reconnection. Most of previous kinetic studies have focused on 
the regime with $\sigma \lesssim 1$, and found a number of
acceleration mechanisms such as direct acceleration at X-line 
regions \citep{Drake2005,Fu2006,Pritchett2006,Oka2010,Huang2010}
and Fermi-type acceleration in reconnection induced plasma flows within magnetic islands 
\citep{Drake2006,Drake2010,Oka2010,Huang2010}. 
The high-$\sigma$ regime ($\sigma > 1$) has been explored 
in a number of papers using the Harris equilibrium 
\citep{Zenitani2001,Zenitani2007,Liu2011,Bessho2012,Cerutti2013,Sironi2014,Melzani2014b,Werner2014}.
However, the initial condition employed in these studies requires 
a hot plasma component inside the current sheet to maintain force
balance, which may not be justified for high-$\sigma$ plasmas. 
Recently, several studies have reported hard power-law
distributions $1 \leq p \leq 2$ when $\sigma \gg 1$
\citep{Sironi2014,Guo2014,Melzani2014b,Werner2014}. 
For a Harris current layer, it was found that a
power-law distribution can be obtained by subtracting the 
initial hot plasma component in the current layer
\citep{Sironi2014,Melzani2014b,Werner2014}. 
In contrast, \citet{Guo2014} used a force-free current sheet 
that does not require the hot plasma population and showed 
that the energy distribution of particles within the entire
reconnection layer develops a power-law  distribution. 
In this study, the primary acceleration mechanism was demonstrated to be
a first-order Fermi mechanism resulting from the curvature drift of particles 
in the direction of the electric field induced by the relativistic flows. 
This mechanism gives rise to the formation of hard power-law spectra 
$f \propto (\gamma-1)^{-p}$ with 
spectral index approaching $p = 1$ for a sufficiently high $\sigma$ and 
a large system size.
An analytical model was developed to describe 
the main feature of the simulations and it gives a general condition 
for the formation of the power-law particle energy distribution. The solution also 
appears to explain simulations from the Harris current layer, 
in which the particles initially in the current layer
form a heated thermal distribution and particles injected from the upstream region 
are accelerated into a power-law distribution
\citep{Sironi2014,Melzani2014b,Werner2014}.


Another important issue is the influence of 3D dynamics that may 
significantly modify the reconnection rate, energy release, and
particle acceleration process. 
Recently, the rate of 3D nonrelativistic magnetic reconnection 
has been explored and compared with 2D simulations in a number of non-relativistic studies \citep{Liu2013,Daughton2014}, which showed only modest differences between
2D and 3D simulations although strong 3D effects emerge as the tearing mode develops over a range of 
oblique angles \citep{Daughton2011}. 
For relativistic magnetic reconnection with a pair plasma, 
\citet{Sironi2014} reported a factor of four
decrease of reconnection rate for 3D simulations compared to 2D simulations. 
This is in contrast to \citet{Guo2014}, who observed
similar reconnection rate and energy conversion between 2D and 3D simulations,
although the kink mode \citep{Daughton1999} strongly 
interacts with the 
tearing mode leading to
a turbulent reconnection layer \citep{Yin2008}.
For particle acceleration in 3D reconnection simulations, early studies reported
that the drift kink instability 
can modify the electric and magnetic field structures in an antiparallel 
reconnection layer and prohibit nonthermal acceleration \citep{Zenitani2005b,Zenitani2007,Zenitani2008}. 
However, recent large-scale 3D
simulations have found strong nonthermal particle spectra 
are produced even when the kink mode is active \citep{Liu2011,Sironi2014,Guo2014}.
Earlier large-scale 3D studies have shown 
the development of turbulence in the reconnection layer \citep{Yin2008}, but its effect to energetic particle acceleration is unknown.
It is therefore important to further study particle
acceleration in relativistic regimes using large-scale
3D kinetic simulations.

In this paper, we perform 2D and 3D fully kinetic simulations 
starting from a force-free current sheet with
uniform plasma density and temperature to
model reconnection over a broad range in the magnetization 
parameter $\sigma = 0.25$ - $1600$. This paper builds upon 
earlier work \citep{Guo2014} and gives further details 
regarding the plasma dynamics and particle acceleration during 
relativistic magnetic reconnection in the high-$\sigma$ regime.
We also present detailed results from a 3D simulation 
that shows a turbulent reconnection layer arising from the interaction between the 
secondary tearing and kink modes.
In Section 2, we describe the numerical methods and 
parameters. 
Section 3 discusses the main results of the paper.
In section 4, we present an analytical model that
explains the main feature of particle acceleration in 
the simulations. The implications from this work for a range of astrophysical problems 
are discussed in Section 5 and our conclusions are summarized in Section 6.
In addition, we have also explicitly examined the numerical convergence for 
this problem and the effect of numerical heating
in our simulations, which is discussed in the Appendix.

\section{Numerical Methods}

We envision a situation where intense current sheets are developed within a magnetically
dominated plasma. Earlier work in non-relativistic low-$\beta$ plasmas has shown that the gradual evolution of the magnetic field
can lead to formation of intense nearly force-free current layers where magnetic reconnection
may be triggered \citep{Titov2003,Galsgaard2003}. In the present study, the critical parameter is the magnetization 
parameter defined as $\sigma \equiv B^2/(4\pi n_e m_e c^2)$, 
which roughly corresponds to the available magnetic energy per particle. The 
numerical simulations
presented in this paper are initialized from a 
force-free current layer with 
$\textbf{B} = B_0 \text{tanh} (z/\lambda) \hat{x} + B_0 \text{sech} (z/\lambda) \hat{y}$
\citep{Che2011,Liu2013,Liu2014}, which 
corresponds to a magnetic field with magnitude $B_0$ 
rotating by $180^\circ$ across the central layer with
a half-thickness of $\lambda$.  No external guide field is 
included in this study but there is an intrinsic guide field $B_y$
associated with the central sheet. The plasma consists of 
electron-positron pairs with mass ratio $m_i/m_e = 1$. The initial distributions 
are Maxwellian with a spatially uniform density $n_0$ 
and a thermal temperature ($kT_{i}=kT_{e}=0.36 m_ec^2$). 
Particles in the central sheet have a net drift $\textbf{U}_i 
= - \textbf{U}_e$ to represent a current density $\textbf{J} = en_0(\textbf{U}_i - \textbf{U}_e)$ that is consistent with 
$\nabla \times \textbf{B} = 4\pi \textbf{J}/c$. Since the force-free
current sheet does not require a hot plasma component to balance the 
Lorentz force, this initial setup is more suitable to study reconnection 
in low $\beta$ and/or high-$\sigma$ plasmas. The full particle
simulations are performed using the VPIC code \citep{Bowers2009} and NPIC code 
\citep{Daughton2006,Daughton2007}, both of 
which solve Maxwell equations and push particles 
using relativistic approaches. The VPIC code directly solves
electric and magnetic fields in Maxwell equations, whereas in the NPIC code the fields are advanced using the scalar and vector potentials.
Therefore the two codes have different algorithms and can be used to cross-check numerical results 
for this problem. We find that they give consistent results for this problem. In addition,
we have developed a particle-tracking module to analyze the detailed physics of the particle energization process. In the simulations, 
we define and adjust $\sigma$ by 
changing the ratio of the electron gyrofrequency 
$\Omega_{ce} = eB/(m_e c)$ to the
electron plasma frequency $\omega_{pe} = \sqrt{4 \pi ne^2/m_e}$, $\sigma \equiv B^2/(4\pi n_e m_e c^2) = (\Omega_{ce}/
\omega_{pe})^2$.
For 2D simulations, we have performed simulations with 
$\sigma = 0.25 \rightarrow 1600$ and box sizes 
$L_x\times L_z =300 d_i \times 
194 d_i$, $600 d_i \times 388 d_i$, and $1200 d_i 
\times 776 d_i$, where $d_i$ is the inertial length $c/\omega_{pe}$. For 3D simulations, the largest case 
is $L_x \times L_y \times L_z = 300 d_i \times 194 
d_i \times 300 d_i$ with 
$\sigma=100$. For high-$\sigma$ cases ($\sigma>25$), we choose 
cell sizes $\Delta x = \Delta y = 
1.46/\sqrt{\sigma} d_i$ and $\Delta z = 0.95/\sqrt{\sigma} d_i$, so 
the particle gyromotion scale $\sim v_{the}/\sqrt{\sigma}d_i$ is resolved. 
The time step is chosen to correspond to a Courant number $C_r = c\Delta t / \Delta r  = 0.7$, where $\Delta r = \Delta x \Delta y \Delta z / (\Delta x \Delta y + \Delta y \Delta z + \Delta x \Delta z)$.
The half-thickness of the current sheet is $\lambda = 6 d_i$ for $\sigma \leq 100$, $12 d_i$ for $\sigma=400$, and $24 d_i$ for $\sigma=1600$ in order to satisfy the drift velocity
$\text{U}_i < c$. 
For both 2D and 3D simulations, we have more than 
$100$ electron-positron pairs in each cell. The boundary conditions for 2D 
simulations are periodic for both fields and particles in the $x$-direction, 
while in the $z$-direction the boundaries are conducting for the field 
and reflecting for the particles.
In the 3D simulations, the boundary conditions are 
periodic for both fields and particles in the 
$y$-direction, while the boundary conditions in the 
$x$ and $z$ directions are the same as the 2D cases.
A weak long-wavelength perturbation \citep{Birn2001} 
with $B_z = 0.03 B_0$ is included to 
initiate reconnection. 
The parameters for different runs are summarized in Table $1$, 
which also lists key results such as maximum energy of particles, 
spectral index, the fraction of kinetic energy converted 
from the magnetic energy and the portion of energy gain arising from the perpendicular electric fields.

Using the set of numerical parameters described above, all of 
the simulations show excellent energy conservation with
violation of energy conservation less than $10^{-3}$ of the
total energy in all cases. 
However, we note that to accurately determine 
the particle energy spectra, the violation in energy conservation
should be smaller than
the initial plasma kinetic energy, which is only a small fraction of the 
total energy for 
the problem we study. Caution is needed when using a small number of particles 
per cell and a small initial plasma kinetic energy in the simulations 
\citep{Sironi2014}, since
numerical heating may significantly modify the particle distribution.
In the Appendix, we have extensively tested how the numerical convergence varies with the initial plasma
temperature, cell size, number of particles per cell, and time step.
For all the cases we present in the main paper,
the violation of energy conservation is a few percent of
the initial kinetic energy in the system, 
meaning effects such as numerical heating have 
a negligible influence on the simulated energy spectra.

\section{Simulation results}

\subsection{General feature and energy conversion}

Figure $1$ gives an overview of the evolution of the 
current layer in 
the case with $\sigma = 100$ and domain size 
$L_x \times L_z = 300d_i \times 194 d_i$ ($L_y=300d_i$ 
for the 3D simulation) from runs 2D-7 and 3D-7. Panel 
(a) shows the color-coded current density
from the 2D simulation and Panel (b) shows a 2D cut of the current density
and a 3D isosurface of plasma density colored by the current 
density from the 3D simulation at $\omega_{pe}t = 175$
 and $\omega_{pe}t = 375$, respectively. 
Starting from the initial perturbation, the 
current sheet 
gradually narrows as the current density is 
concentrated in the central region. 
In the 2D simulation, the extended thin current sheet 
 breaks into 
a number of fast-moving secondary plasmoids ($\omega_{pe}t \sim 225$) due to the secondary tearing 
instability. The plasmoids coalesce and eventually merge into a 
single island at the edge of the simulation domain similar to the 
nonrelativistic case \citep{Daughton2007}. 
In the 3D simulation, as the intrinsic guide field
associated with the force-free current layer is expelled from 
the central region, the kink instability
\citep{Daughton1999} develops and interacts with the tearing 
mode, leading to a turbulent evolution \citep{Yin2008}.
However, despite the strong 3D effects that modify the current layer,
small-scale flux-rope-like structures with intense current density 
develop repeatedly as a result of the secondary tearing 
instability. 

\begin{figure*}
\begin{center}
\includegraphics[width=\textwidth]{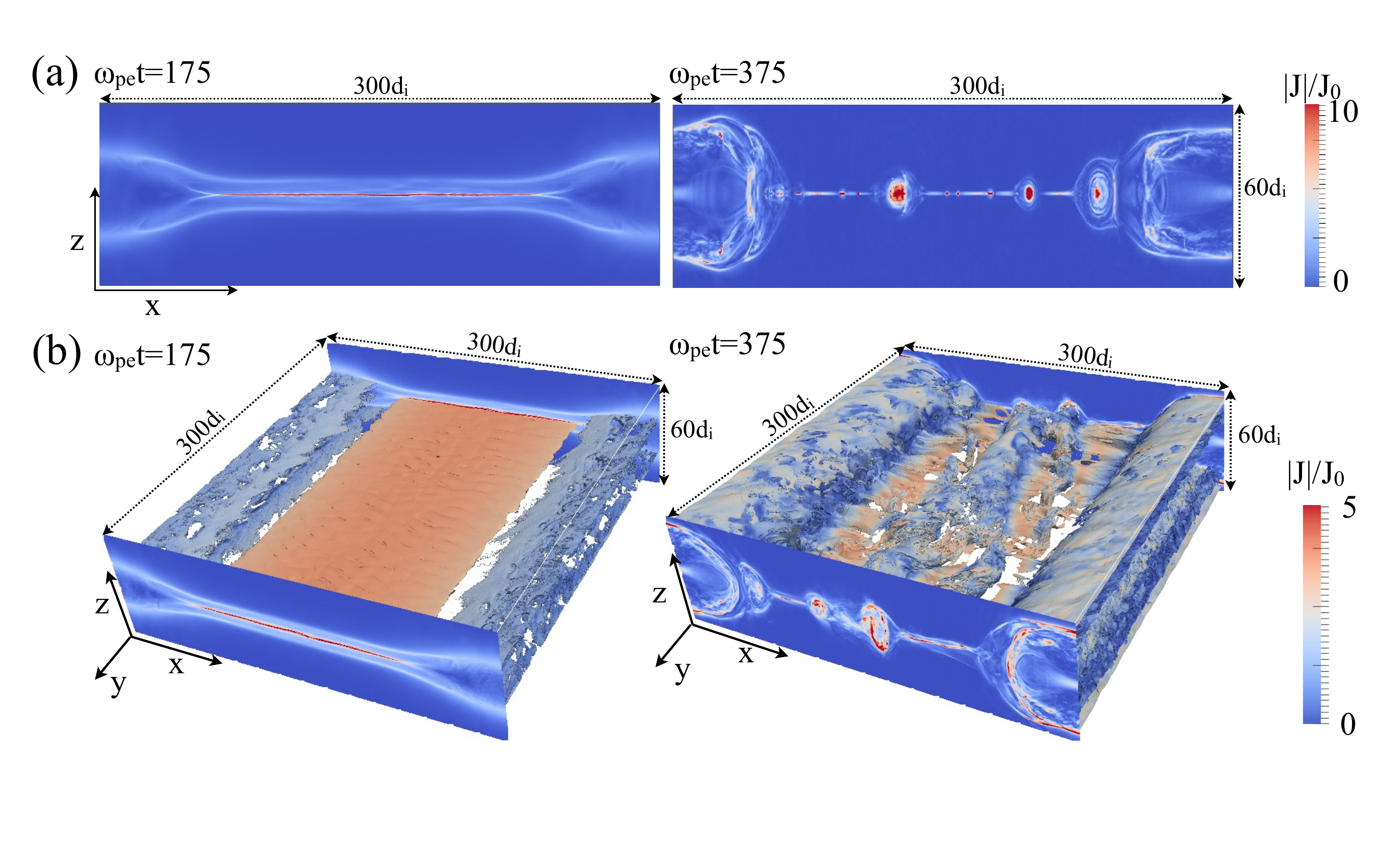}
\caption{Evolution of 2D and 3D simulations with $\sigma =100$ and domain size $L_x \times L_z = 300 d_i \times 194 d_i$ ($L_y = 300 d_i$ for 3D); (a) Color-coded current density from the 2D simulation at $\omega_{pe}t = 175$ and $\omega_{pe}t = 375$, respectively; (b) 2D cut of current density 
and a 3D isosurface of  the
plasma density colored by the current density at $\omega_{pe}t = 175$ and $\omega_{pe}t = 375$, respectively.}
\end{center}
\end{figure*}

Although the plasma dynamics in the 2D and 3D 
simulations appears quite different, the energy conversion and 
particle energization are very similar. Figure 2 (a) shows the evolution of 
magnetic energy $E_B$, electric field energy $E_E$, kinetic energy $E_k$, and energy 
carried by relativistic particles with $\gamma > 4$ 
from the 2D and 3D simulations
(2D-7 and 3D-7). Note in both of these simulations, 
the total energy is conserved to within $10^{-4}$ 
of the initial value. The evolutions of different
forms of 
energies between 2D and 3D simulations are 
very similar. 
In both the 2D and 3D simulations, about $25 \%$ of the 
magnetic energy is converted into 
plasma kinetic energy, most of which is carried 
by relativistic particles. Figure 2 (b) shows the time-integrated energy 
conversion from magnetic energy into plasma energy in the simulation
$\int^t_0 dt \int dV \textbf{J} \cdot \textbf{E}$ and its
contribution from parallel and perpendicular electric field terms
$\textbf{J}_\parallel \cdot \textbf{E}_\parallel$ and $\textbf{J}_\perp \cdot \textbf{E}_\perp$, respectively. Here $\int dV$ = $\int dxdydz$. 
The difference in energy conversion between the 2D and 3D simulations can be as large as a factor of two at $\omega_{pe}t = 300$, but at the end of the 
simulations both cases have converted about the same amount of magnetic energy. This shows that the kink instability that may  modify the magnetic field does
not significantly change the overall energy conversion.
While the energy conversion through 
parallel electric field
is important when the thin current layer initially develops, 
most of the energy conversion is due to perpendicular
electric fields induced by relativistic flows as the system is dominated
by secondary plasmoids/flux ropes. This analysis has been done in all the cases and summarized
in Table 1, which shows that in most of cases, 
the perpendicular electric field plays a 
dominant role in converting magnetic energy 
into plasma kinetic energy. 
This can also be seen in Figure $3$, which shows the color-coded intensities of 
$\textbf{J} \cdot \textbf{E}$, $\textbf{J}_\perp \cdot \textbf{E}_\perp$ 
and $\textbf{J}_\parallel \cdot \textbf{E}_\parallel$ 
from the 2D and 3D simulations at $\omega_{pe}t = 175$ and 
$\omega_{pe}t = 375$, respectively. 
Figure 2 (c) 
compares the energy spectra from the 2D and 3D simulations at various 
times. The most striking feature is that a hard power-law 
spectrum $f \propto (\gamma-1)^{-p}$ with a spectral index $p \sim 1.35$ 
forms in both 2D and 3D runs. 
Although a fraction of particles are accelerated in the early phase when the
parallel electric field is important, most of the particles in the power-law 
distribution are accelerated when the system is dominated by plasmoids/flux ropes.
As we will discuss below, the formation of power law is closely related to 
the motional electric field induced by the fast moving plasmoids.
In the subpanel, the energy spectrum for all 
particles in the 3D simulation at $\omega_{pe}t=700$ is shown 
by the red line. The low-energy portion can be fitted by a Maxwellian 
distribution (black) and the nonthermal part 
resembles a power-law distribution (blue) starting at $\gamma \sim 2$ with an exponential cut-off for $\gamma \gtrsim 100$. The nonthermal 
part contains $\sim25\%$ of particles and $\sim95\%$ of the
kinetic energy. The maximum particle energy of the system
can be predicted approximately using the
reconnecting electric field 
$m_ec^2(\gamma_{max}-1) = \int |qE_{rec}|c dt$ until 
the gyroradius is comparable to the system size (see also Figure 6b). 
Although we observe a strong kink instability in the 3D simulations, the energy conversion and particle 
energy spectra are remarkably similar to the 2D results, indicating the 3D effects 
are not crucial for the particle acceleration. The fast 
acceleration is distinct from that of nonrelativistic magnetic reconnection,
where particles are at most accelerated to mildly relativistic energy \citep[e.g.,][]{Fu2006,Drake2006,Pritchett2006,Oka2010}. 
The nonthermal-dominated distribution in the simulations is also quite different from distributions  
in the relativistic shock regions
\citep[e.g.,][]{Spitkovsky2008}, where the particles are heated at the shock front 
and form an extended thermal distribution containing most of the dissipated energy. 
The power-law spectral index $p \sim 1$ from relativistic reconnection
is significantly harder than the limit $p \sim 2$ predicted by nonrelativistic
and relativistic shock acceleration theories \citep[e.g.,][]{Blandford1987,Achterberg2001}.

\begin{figure}
\begin{center}
\includegraphics[width=0.5\textwidth]{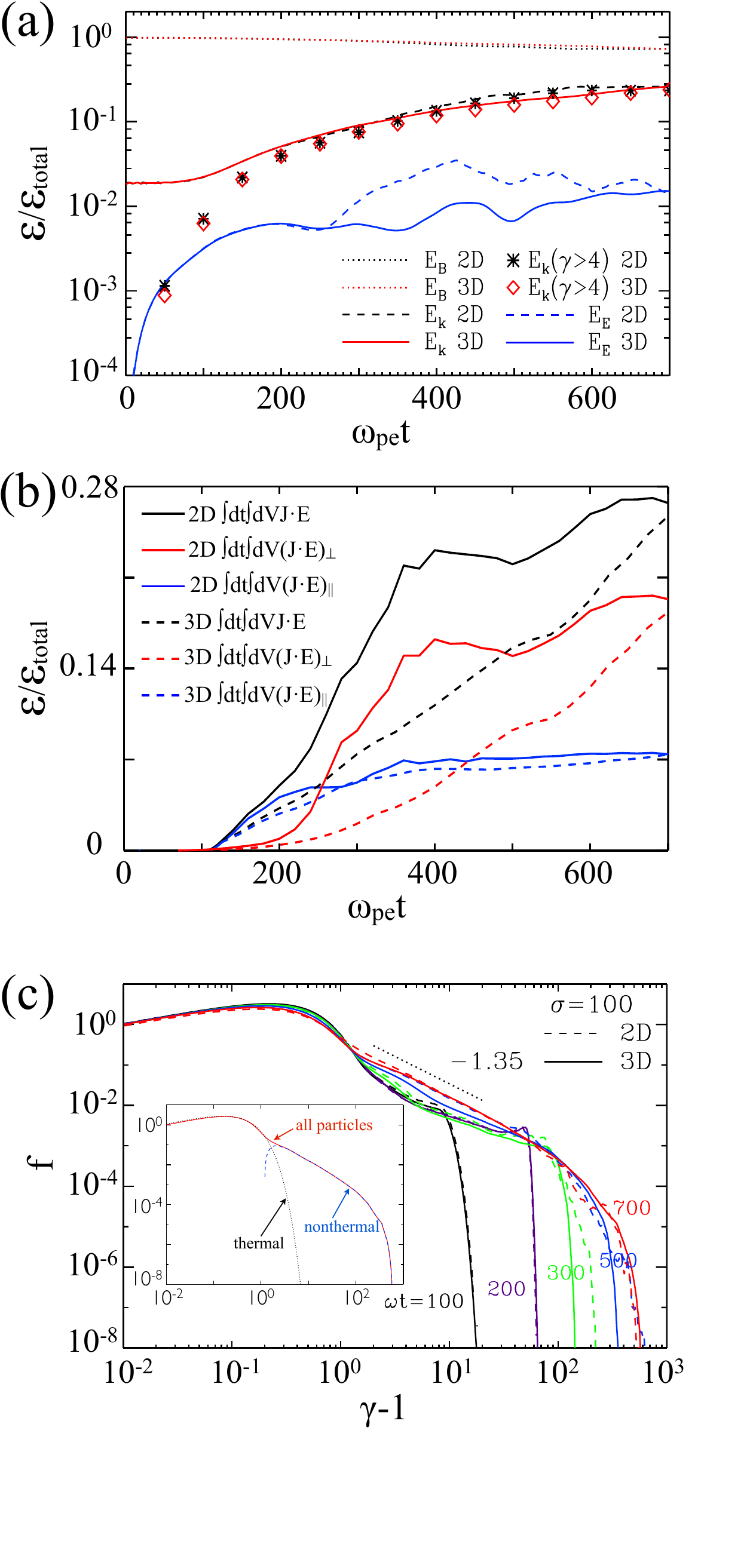}
\caption{Plasma energetics in 2D and 3D simulations 
with $\sigma =100$ and domain size 
$L_x \times L_z = 300 d_i \times 194 d_i$ 
($L_y = 300 d_i$ for 3D); (a) Evolution of 
magnetic energy $E_B$, electric field energy 
$E_E$, plasma 
kinetic energy $E_k$ and energy carried by 
relativistic particles with Lorentz factor 
$\gamma >4$; (b) Energy conversion from magnetic 
energy into plasma energy 
integrated over time 
$\int^t_0 dt \int dV\textbf{J} \cdot \textbf{E}$ 
and its
contribution from parallel and 
perpendicular electric field 
$\textbf{J}_\parallel \cdot \textbf{E}_\parallel$ and 
$\textbf{J}_\perp \cdot \textbf{E}_\perp$;
(c) Evolution of particle
energy spectra from 2D and 3D simulations. Subpanel: 
Energy spectrum from the 3D 
simulations at $\omega_{pe}t = 700$. The low energy is 
fitted with a thermal distribution and rest of the 
distribution is a nonthermal power law with an 
exponential cutoff.}
\end{center}
\end{figure}

\begin{figure*}
\begin{center}
\includegraphics[width=\textwidth]{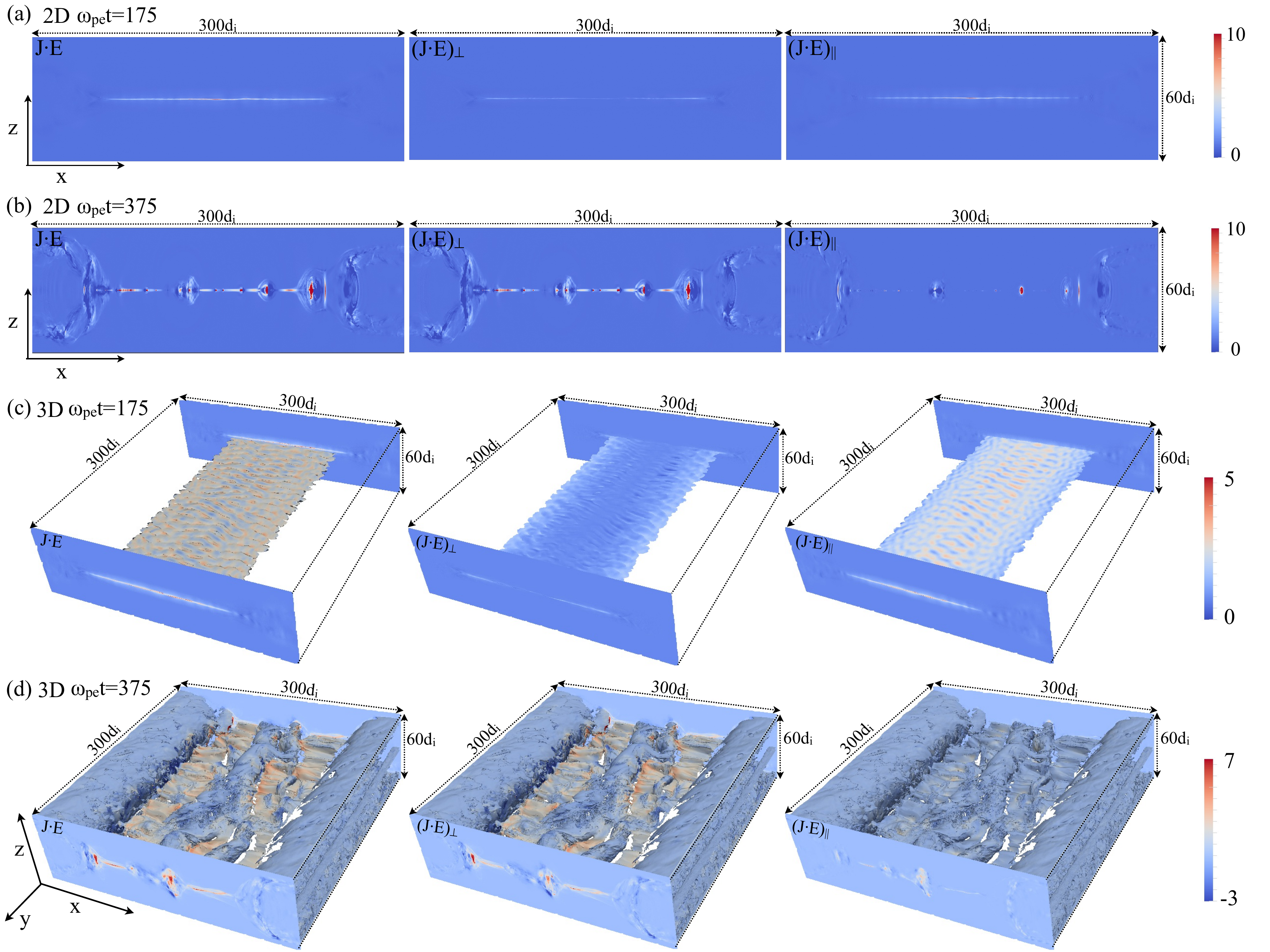}
\caption{Color-coded intensity of energy conversion
rate $\textbf{J} \cdot \textbf{E}$ normalized using
$n_0 m_e c^2 \omega_{pe}$ and contributions
from $\textbf{J}_\perp \cdot \textbf{E}_\perp$ 
and $\textbf{J}_\parallel \cdot \textbf{E}_\parallel$ 
for the 2D
and 3D simulations with $\sigma = 100$ at 
$\omega_{pe}t = 175$ and 
$\omega_{pe}t = 375$, respectively. 
In the early stage
the conversion by parallel electric 
field is important and the 
perpendicular electric field 
plays a dominant role when multiple-plasmoids (flux ropes in 3D) 
develop due to the secondary tearing instability.}
\end{center}
\end{figure*}

\subsection{Particle Acceleration}
We now discuss the details of particle acceleration. We will first
present some analysis of particle trajectories to show the acceleration mechanism. 
Then the dominant acceleration mechanism is distinguished by
tracking all the particles and calculating the 
energy gain using the guiding-center drift 
approximation. The results demonstrate that
the dominant acceleration mechanism is a 
first-order Fermi acceleration through curvature drift motion 
along the motional electric field induced by the 
relativistic reconnection flows. 
We calculate the acceleration rate $\alpha = \Delta \varepsilon/(\varepsilon \Delta t) $ and its time 
integral for cases with $\sigma = 6 - 400$, where $\Delta \varepsilon$ is the averaged energy gain  for particles of energy 
$\varepsilon$ over a period $\Delta t$. Finally, we 
summarize the character of the energy spectra. These main results will 
be discussed and interpreted in detail in Section 4, where we present the acceleration model.

Figure 4 and Figure 5 present the trajectory analysis for the motions of accelerated particles
in the 2D case with $\sigma = 100$ and $L_x \times L_z = 600 d_i \times 388 d_i$. These particles are selected to show the
common acceleration pattern of accelerated particles, which is consistent with the results of statistical analysis for the acceleration of particles in Figure 6.
The first three panels of Figure 4 show (a) the trajectory of a representative particle close to the central sheet between 
$\omega_{pe}t = 30$ - $300$ together with $E_\parallel$ at 
$\omega_{pe}t = 180$, (b) the trajectory of the same particle between 
$\omega_{pe}t = 310$ - $510$ together with $E_y$ at $\omega_{pe}t = 400$,
and (c) the trajectory of the particle between 
$\omega_{pe}t = 510$ - $720$ together with
$E_y$ at $\omega_{pe}t = 640$, respectively. The starting and ending locations of the particle 
are labeled by `$+$' and `$\times$' signs, respectively. 
Note that the field is highly variable in time 
and the location of the particle at the same time 
step as the field contour is drawn by the
`$\ast$' sign. 
The two bottom panels
show the evolution of the particle energy as a 
function of time (d) and energy as a function of the 
$x$ position (e), respectively.
Each period corresponding to that in 
($a$)-($c$) is labeled
by the same color. 
The green curve represents
the energy gain in the parallel electric field 
integrated from $t = 0$.
Initially the particle is close to the central layer
and gains energy by the parallel electric field.
It is then strongly accelerated by perpendicular electric field
when the reconnection region breaks into multiple islands and
the electric field is mostly the motional 
electric field $\textbf{E} =
-\textbf{V} \times \textbf{B} /c$ generated by 
relativistic plasma outflows. The figure also shows
that the acceleration by $\textbf{E}_\perp$ resembles a Fermi
process by bouncing back and forth within a magnetic 
island.

\begin{figure*}
\begin{center}
\includegraphics[width=0.7\textwidth]{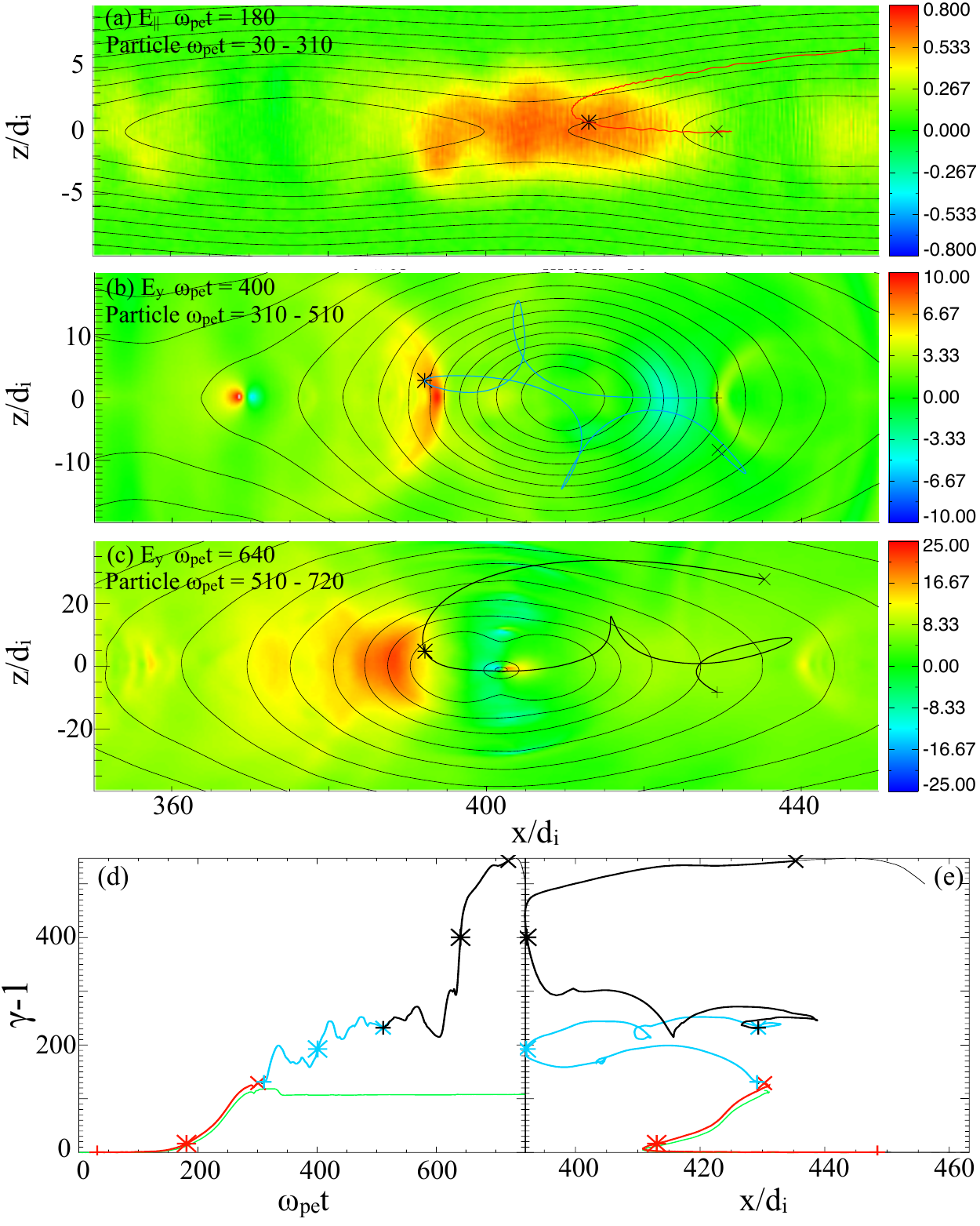}
\caption{Panels (a)-(c) show a particle trajectory in the $x$-$z$ plane together with
the color-coded electric field (a) $E_\parallel$, (b) $E_y$, and (c) $E_y$. 
Panels (d) and (e) show the particle energy 
as a function of time and energy as a function of the $x$ position, respectively. In (d) and (e), 
curves with different colors represent the energy evolution during time periods in (a)-(c). The green curve shows the integrated
energy gain from the parallel electric field.}
\end{center}
\end{figure*}

Figure $5$ presents another view of the particle acceleration physics. 
It is similar to Figure 4, but the field contours show 
the outflow speed to highlight the role of $V_x$ in the
particle's energization. This clearly illustrates a relativistic 
first-order Fermi process by bouncing in outflow regions of the 
reconnection layer. Note the energy gain from the parallel electric 
field for this sample particle is negligible since it entered the 
reconnection layer longer after the development of multiple plasmoids.

\begin{figure*}
\begin{center}
\includegraphics[width=0.7\textwidth]{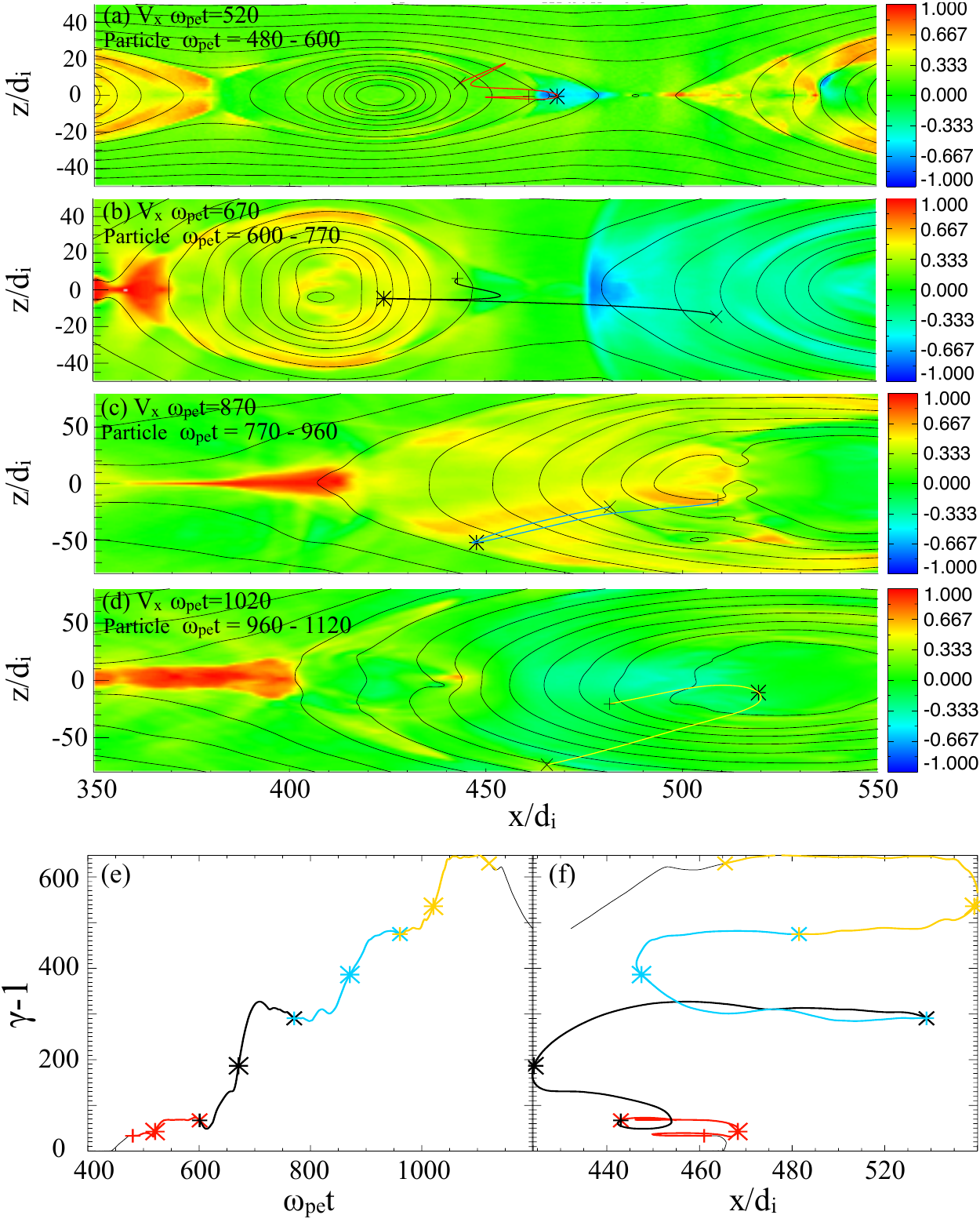}
\caption{Panels (a)-(d) show a particle trajectory in the $x$-$z$ 
plane together with the fluid velocity in the $x$ direction $V_x$.
Panels (e) and (f) show the particle energy 
as a function of time and energy as a function of the $x$ position, 
respectively. Different colored
curves represent the energy evolution during 
time periods in (a)-(d), showing that the particle gains energy 
by bouncing in the relativistic flow generated by reconnection.
}
\end{center}
\end{figure*}

In Figure 6, we present more analysis for the 
mechanism of particle acceleration. 
Panel (a) shows the energy as a 
function of the $x$-position of four accelerated particles.  Similar to Figure 5, the electrons gain energy by bouncing back and forth within the reconnection layer. 
We have analyzed trajectories of a large number of particles and 
found the energy gain for each cycle is $\Delta \varepsilon \sim \varepsilon$, 
which demonstrates that 
the acceleration mechanism is a first-order Fermi process 
\citep{Drake2006,Drake2010,Kowal2011}. Panel (b) shows the 
maximum particle energy in the system as a function 
of time. This is plotted using different count level from the 1-particle level 
to the 1000-particle level. Also plotted is the estimated maximum energy resulting
from the reconnecting electric field by assuming particles moving along the
electric field at the speed of light $\int |qE_{rec}|c dt$. This shows that
the maximum possible energy occurs for a small number of particles that continuously 
sample the reconnection electric field 
$m_ec^2\gamma_{max} = \int |qE_{rec}|c dt$. At late time, as the particle
gyroradius becomes large and comparable to the system 
size, the maximum energy saturates.
To show
the Fermi process more rigorously, we have 
tracked the energy change 
for all the particles in the simulation and the relative contributions arising
from the parallel electric field 
($m_ec^2\Delta \gamma = \int qv_{\parallel} E_\parallel dt$) 
and curvature drift 
acceleration ($m_ec^2 \Delta \gamma = \int q \textbf{v}_{curv} \cdot \textbf{E}_{\perp} dt$) similar to \citep{Dahlin2014}, 
where $\textbf{v}_{curv} = \gamma v^2_\parallel (\textbf{b} \times (\textbf{b}\cdot \nabla)\textbf{b})/\Omega_{ce}$, $v_\parallel$ is the particle velocity parallel to the magnetic field, and $\textbf{b} = \textbf{B} / |B|$.
Panel (c) shows the averaged energy gain and the contribution from parallel 
electric field and curvature drift acceleration over an 
interval of $25 \omega_{pe}^{-1}$ as a function of energy starting at $\omega_{pe}t=350$. 
The energy gain follows $\Delta \varepsilon \sim \alpha \varepsilon$, confirming the first-order 
Fermi process identified from particle trajectories. The energy gain 
from the parallel motion depends weakly on energy, whereas the energy gain from the
curvature drift acceleration is roughly proportional to energy. In the early phase, the parallel electric field is strong but only 
accelerates a small portion of particles, and the curvature drift  dominates the 
acceleration starting at about $\omega_{pe}t = 250$. 
The contribution from the gradient drift was also evaluated and found to be negligible in comparison.
Panel (d) shows  $\alpha = 
<\Delta \varepsilon>/(\varepsilon \Delta t)$ measured directly from the energy gain of the particles in the perpendicular electric field ($m_ec^2 \Delta 
\gamma = \int q \textbf{v}_{\perp} \cdot \textbf{E}_{\perp} dt$) and estimated from the 
expression for the curvature drift acceleration.
The close agreement demonstrates that curvature drift 
term dominates the particle energization. 

\begin{figure*}
\begin{center}
\includegraphics[width=0.7\textwidth]{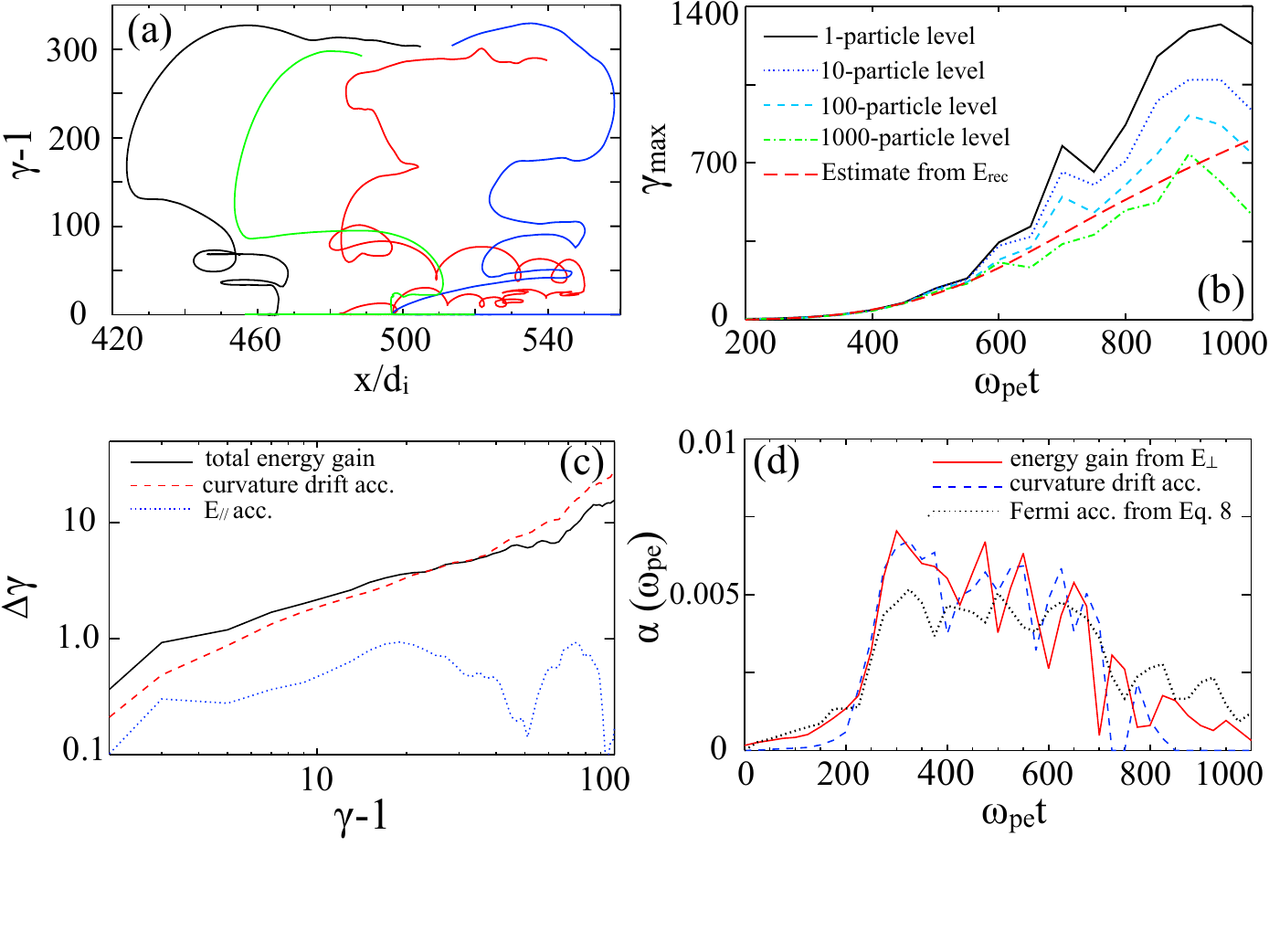}
\caption{(a) Energy as a function of $x$-position of four
accelerated particles; (b) The maximum energy of particles 
in the system as a function of time from the $1$-particle
count level to the $1000$-particle count level. The red dashed 
line shows the maximum energy estimated for a particle moving in the direction of the reconnecting electric field at the speed 
of light $m_ec^2\gamma_{max} = \int |qE_{rec}|c dt$; (c) 
Averaged energy gain and contributions from parallel 
electric fields and curvature drift acceleration over 
a time interval of 25$\omega_{pe}^{-1}$ as a function of 
particle energy starting at $\omega_{pe}t=350$;(d) $\alpha = 
<\Delta \varepsilon>/(\varepsilon \Delta t)$ from energy gain in 
perpendicular electric field and by curvature drift acceleration, 
and from the Equation (6) using the averaged flow speed and 
island size.}
\end{center}
\end{figure*}

For higher $\sigma$ and 
larger domains, the acceleration is stronger and 
reconnection is sustained over a longer 
duration. 
In Figure 7(a), we present the energy spectra at the
end of simulation for a number of cases with different $\sigma$ and system size $L_x \times L_z = 600 d_i \times 388 d_i$.
A summary for the spectral index can be found in 
Table $1$.
In Figure 7(b), a summary for the measured spectral 
index for the power-law ranges of all the 2D runs 
shows that the spectrum is 
harder for higher $\sigma$ and larger domain sizes, 
and approaches the limit $p=1$. Note that the 
spectral indexes appear systematically harder than in
other recent papers \citep{Sironi2014,Melzani2014b,Werner2014}. However,
the energy spectra in these studies are plotted
using total relativistic energy $\gamma mc^2$ and
here we use kinetic energy $(\gamma-1) mc^2$. Using 
total relativistic
energy in the energy spectra significantly distorts
the spectral index in the energy range of $0 < \gamma - 1 < 10$, which may alter 
the interpretation of the results 
\citep{Sironi2014,Melzani2014b,Werner2014}\footnote{In fact, our simulation results show that 
the ``$-1$'' spectra can be obtained as
long as the magnetic energy dominates over the initial plasma 
kinetic energy $8 \pi nkT_0/B^2= \beta \ll 1$. An example can 
be seen in the Appendix (Figure 13), which robustly shows the 
$p=1$ spectrum can be obtained when $\sigma=25$ and 
$kT_0 = 0.01 mc^2$. The same spectrum gives a ``$p \sim 2$'' 
slope when it is plotted as a function of $\gamma$, which may 
explain the different conclusions reported
by other papers \citep{Sironi2014,Melzani2014b,Werner2014}}.

\begin{figure}
\begin{center}
\includegraphics[width=0.42\textwidth]{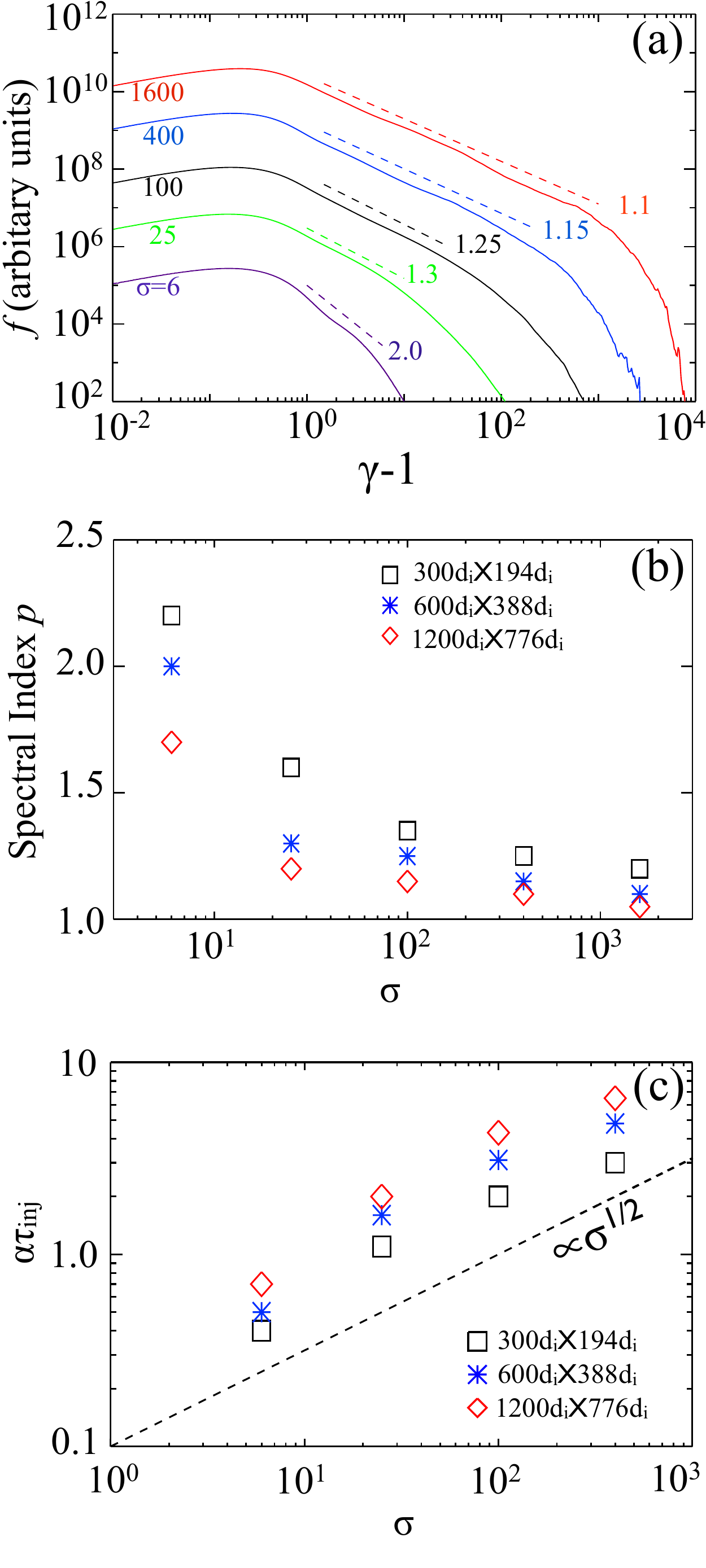}
\caption{Energy spectra at the end of 
simulations for a 
series of 2D runs with system size $L_x \times L_z = 600 d_i \times 388 d_i$ and different $\sigma$ from 
6 to 1600. (b) Spectra index for all 2D simulations with $\sigma$ from 6 to 1600. (c) Time integrated $\alpha \tau_{inj}$ for cases with $\sigma = 6$-$400$ and different system sizes.}
\end{center}
\end{figure}

\subsection{Reconnection rate and relativistic flows}

Figure 8 (a) shows the time-dependent reconnection rates
normalized using the initial asymptotic magnetic field $B_0$ 
in 2D and 3D simulation with $\sigma = 100$ (Run 2D-7 
and 3D-7). The 2D reconnection rate is computed from 
\begin{eqnarray}
R = \frac{E_{rec}}{B_0}= \frac{1}{B_0V_{A0}}<\frac{\partial \psi}{\partial t}>, \nonumber
\end{eqnarray}
where $\psi = \max(A_y) - \min(A_y)$ along the central layer $z=0$, $A_y$ is the vector potential along the $y$ direction, $<>$ represents a time average over $\delta t \omega_{pe} = 25$ \citep{Liu2014}, $V_{A0} = v_A/\sqrt{1+(v_A/c)^2} =  \sqrt{\sigma/(2+\sigma)}c$ is the relativistic Alfven speed in the cold-plasma limit. 
Here $v_A = B_0/\sqrt{4 \pi n (m_i + m_e)}$ is the non-relativistic Alfven speed based on $B_0$. 
The 3D reconnection rate is estimated by using the mixing 
of plasma across the separatrix surfaces \citep{Daughton2014}.
The rate in the 2D simulation is quite variable but the range is
within a factor of two times of the 3D results, meaning that the 2D and 
3D simulations give roughly the same reconnection rate. 
Figure 8 (b) shows the peak reconnection rate 
for a number of 2D cases with $\sigma$ from $0.25$ to $1600$ and box size $1200d_i \times 776 d_i$.
The rate is observed to increase with $\sigma$ from 
$E_{rec} \sim 0.03B_0$ for $\sigma=1$ to 
$E_{rec} \sim 0.24B_0$ for $\sigma=1600$.
 It shows that the peak reconnection field increases with 
$\sigma$ and starts to saturate around $\sigma=1000$. 
For low-$\sigma$
cases with $\sigma < 1$, the reconnecting electric field is 
consistent with previous work for nonrelativistic 
reconnection \citep[e.g.,][]{Daughton2007}. 
More detailed analyses have shown that for high-$\sigma$ cases, 
the reconnection rate normalized
using the magnetic field $B_u$ upstream of the diffusion region $E_{rec}/B_u$
is close to $1$ for $\sigma \gtrsim 100$ \citep{Liu2015}.

\begin{figure}
\begin{center}
\includegraphics[width=0.42\textwidth]{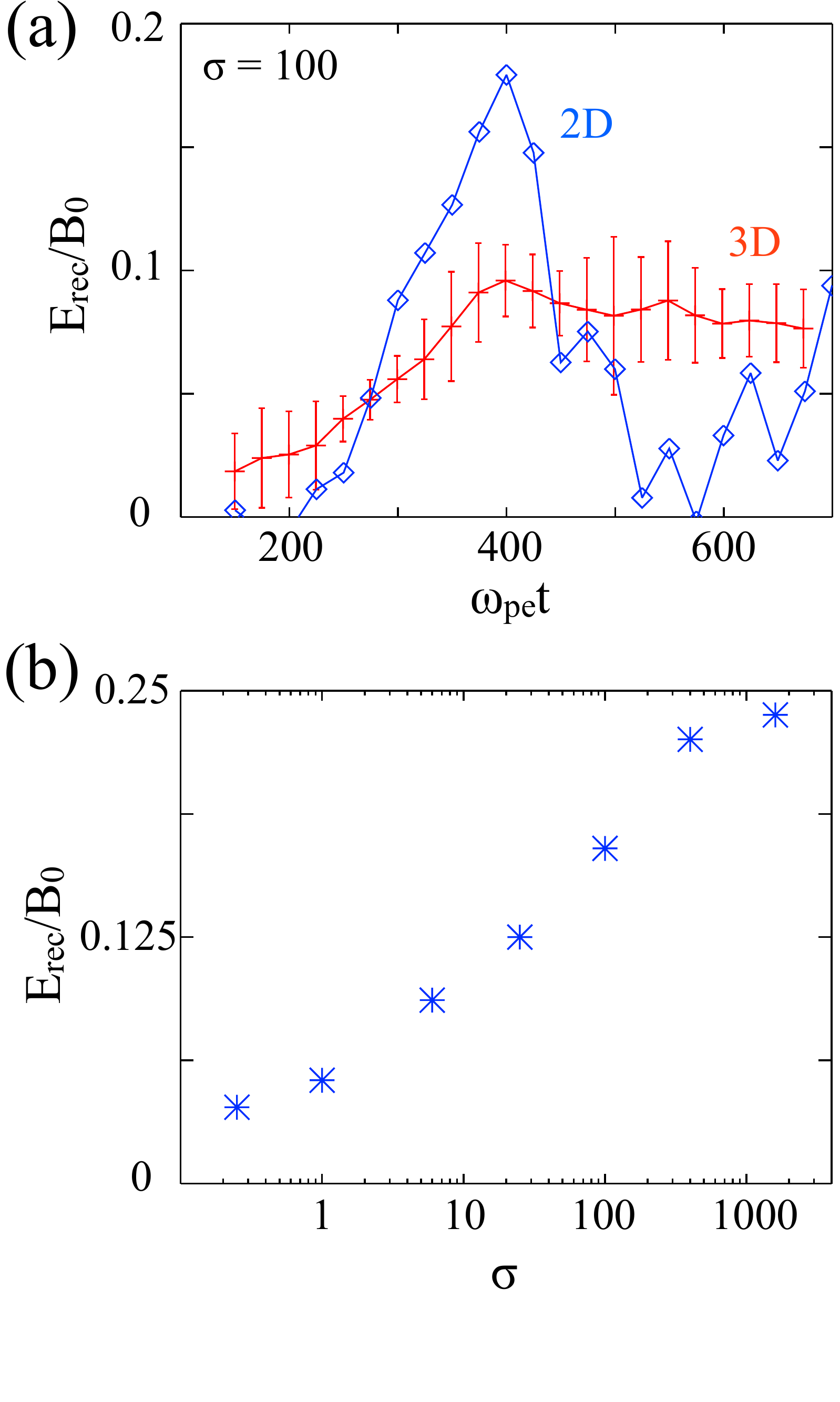}
\caption{(a) Time-dependent 2D and 3D reconnection electric field normalized by the initial magnetic field $E_{rec}/B_0$. (b) Normalized peak electric field $E_{rec}/B_0$ as a function of $\sigma$ in 2D simulations. }
\end{center}
\end{figure}

In Figure 9 (a) and (b) we plot the maximum flow velocity in the $x$
direction (outflow direction) and the corresponding Lorentz factor $\Gamma_x$. The 2D results are represented by blue symbols and
the 3D results are in red symbols, respectively.
Although we have only used a small simulation domain
that may be affected by counter-streaming particles, a
relativistic outflow still develops with $\Gamma_x$ of a few.
In Figure 9 (c) and (d) we plot the maximum flow velocity in the $z$
direction (inflow direction) and the corresponding Lorentz factor $\Gamma_z$, 
respectively. Interestingly, the inflow speed can also be relativistic
for high-$\sigma$ cases.
Detailed analysis for the diffusion region has been
discussed in \citet{Liu2015}, which shows that the inflow speed can
be predicted by a model based on the Lorentz contraction of the plasma passing
through the diffusion region. 

\begin{figure*}
\begin{center}
\includegraphics[width=0.8\textwidth]{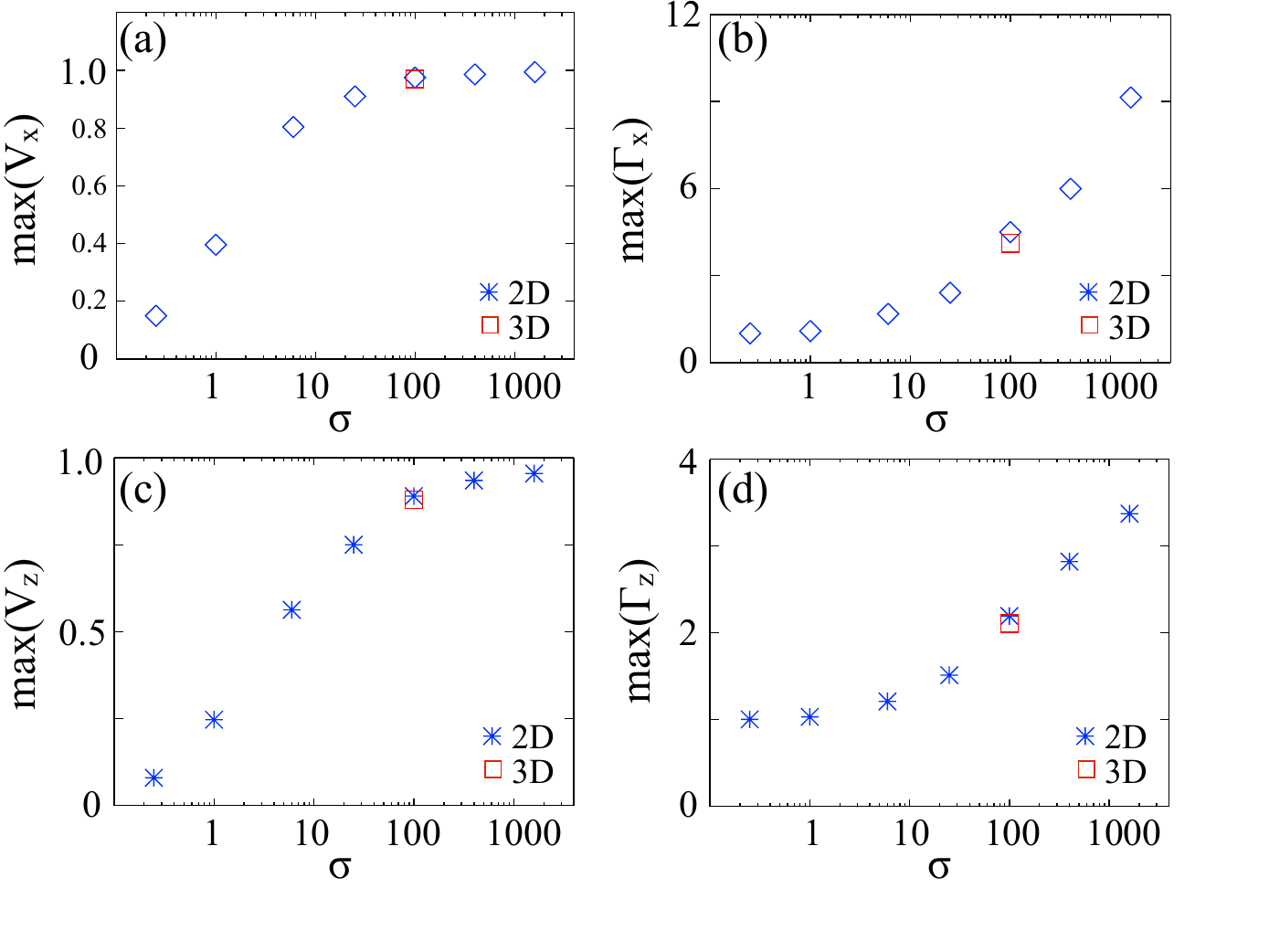}
\caption{(a) The maximum flow velocity in the $x$ direction $V_x$ as a function of $\sigma$; 
(b) The maximum flow Lorentz factor in the $x$ direction $\Gamma_x = 1/(1 - V_x^2/c^2)$
as a function of $\sigma$;
(c) The maximum flow velocity in the $z$ direction $V_z$ as a function of $\sigma$;
(d) The maximum flow Lorentz factor in the $z$ direction $\Gamma_z = 1/(1 - V_z^2/c^2)$
as a function of $\sigma$.}
\end{center}
\end{figure*}

The enhanced reconnection rate and development of relativistic 
inflow/outflow structures are in contrast to the results reported 
earlier \citep{Sironi2014}, where the outflow 
can only be mildly relativistic and the inflow speed
remains nonrelativistic. Note that \citet{Liu2015} has
also reported the development of relativistic inflow
for both Harris and force-free current sheets, indicating that this
property of relativistic magnetic reconnection does
not strongly depend on the initial setup.

\subsection{3D Dynamics}
In our three-dimensional 
simulation, we also find strong bulk $\Gamma_x \sim 4$ can 
develop in the system, meaning the development of relativistic
flows is not strongly influenced by 3D effects.
Figure 10 shows the power spectrum of magnetic fluctuations 
with wave numbers perpendicular to the $y$ direction and a 
volume rendering of the current density in the 3D simulation 
with $\sigma = 100$ at $\omega_{pe}t = 708$. The power 
spectrum shows a clear inertial range with a slope of 
``$-2$'' and steeper slope for higher wave numbers 
$k_\perp d_i \gtrsim 1$. As we have discussed, the 3D 
simulation allows the development and interaction of 
secondary tearing instability and kink instability,
leading to a turbulent magnetic field in the reconnection 
layer. For the 
whole simulation, the range of scales for the 2D magnetic 
islands is similar to the observed 3D flux ropes. The maximum 
energy in both 2D and 3D agrees 
well with the time integral of energy gain from reconnecting 
electric field. This is in contrast to the earlier kinetic 
simulations \citep{Zenitani2005,Zenitani2007,Zenitani2008}.
The energy distributions reported in this paper 
are remarkably similar in 2D and 3D, suggesting that
the underlying Fermi acceleration is rather robust and
does not depend on the existence of well-defined magnetic islands.

\begin{figure*}
\begin{center}
\includegraphics[width=\textwidth]{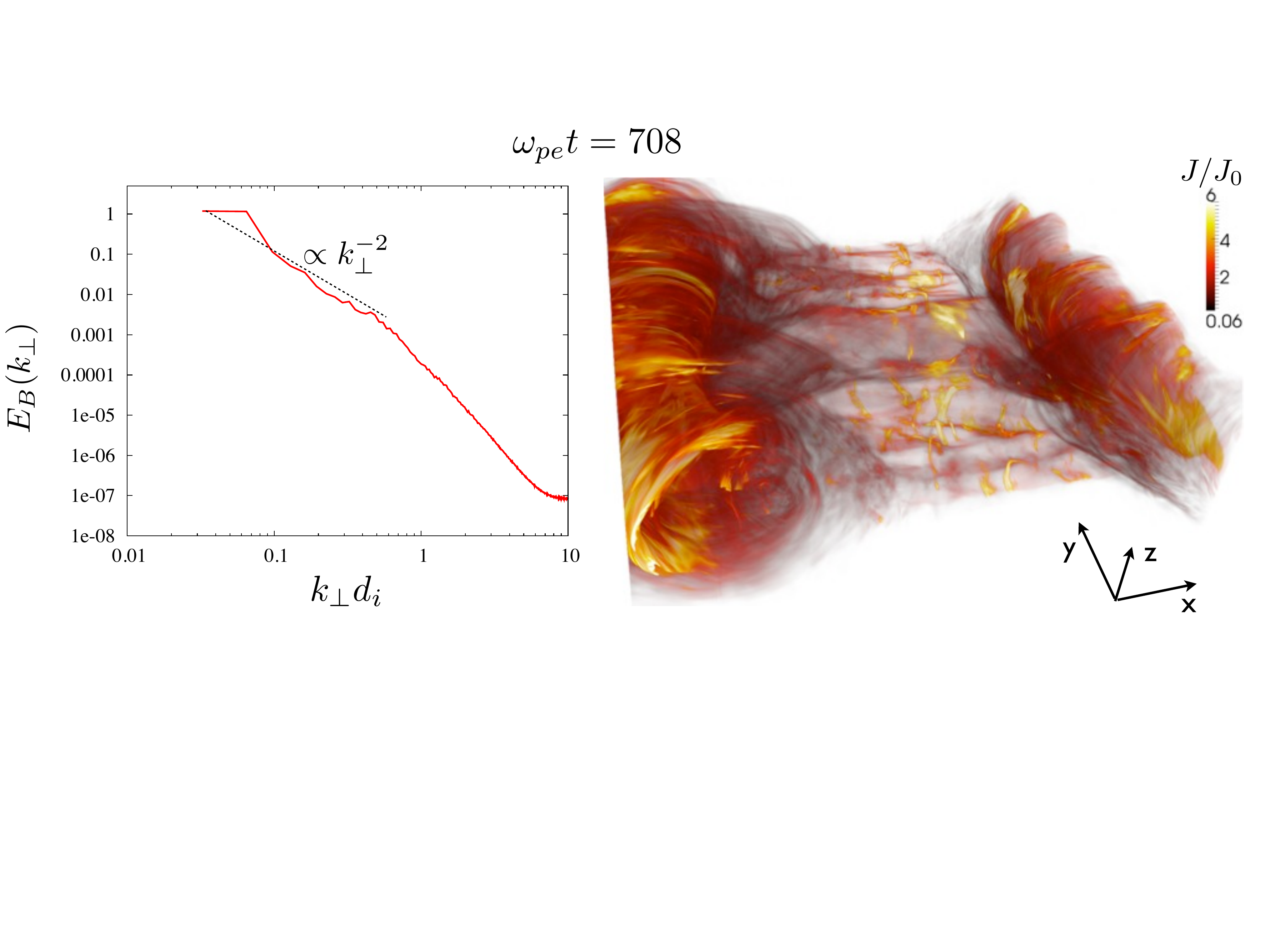}
\caption{The evidence for turbulence in the 3D simulation. Left: 
power spectrum of magnetic fluctuations with wave numbers
perpendicular to the $y$ direction. Right: volume rendering 
of the current density $J/J_0$ in the 3D simulation at $\omega_{pe}t = 708$.}
\end{center}
\end{figure*}

\begin{table*}
\begin{center}
\begin{tabular*}{0.7\textwidth}{ccccccccc}
\hline 
Run   & $\sigma$ & system size & $\lambda$ & p & $\gamma_{max}$  & $E_{kin}\%$ & $(J\cdot E)_\perp\%$  & $\alpha \tau_{inj}$ \\
\hline
2D-1  & 6    & $300 d_i\times 194 d_i$  & $6d_i$   & 2.2  & 45  & 23\% & 83\% & 0.4\\  
2D-2  & 6    & $600 d_i\times 388 d_i$  & $6d_i$   & 2.0  & 56 & 32\% & 92\% & 0.5\\
2D-3  & 6    & $1200 d_i \times 776 d_i$ & $6d_i$   & 1.7  & 79 & 34\% & 93\% & 0.7\\
2D-4  & 25   & $300 d_i \times 194 d_i$  & $6d_i$   & 1.6  & 195 & 28\% & 85\% & 1.1\\  
2D-5  & 25   & $600 d_i \times 388 d_i$  & $6d_i$   & 1.3  & 339 & 37\% & 90\% & 1.6\\
2D-6  & 25   & $1200 d_i \times 776 d_i$ & $6d_i$   & 1.2  & 617 & 42\% & 90\% & 2.0\\
2D-7  & 100  & $300 d_i \times 194 d_i$  & $6d_i$   & 1.35 & 650 & 29\% & $73\%$ & 2.0\\  
3D-7&100&$300 d_i\times 194 d_i\times 300 d_i$& $6d_i$& 1.35 & 617 & 28\% &$71\%$ & N/A\\  
2D-8  & 100  & $600 d_i\times 388 d_i$  &    $6d_i$& 1.25 & 1148 & 40\% &$78\%$ & 3.1\\
2D-9  & 100  & $1200 d_i\times 776 d_i$ &    $6d_i$& 1.15 & 1862 & 45\% &$94\%$ & 4.3\\
2D-10 & 400  & $300 d_i\times 194 d_i$  &   $12d_i$& 1.25 & 1514 & 20\% &$54\%$ & 3.0\\  
2D-11 & 400  & $600 d_i\times 388 d_i$  &   $12d_i$& 1.15 & 3715 & 31\% &$75\%$ & 4.8\\
2D-12 & 400  & $1200 d_i \times 776 d_i$ &   $12d_i$& 1.1  & 5495 & 36\% &$86\%$ & 6.5\\
2D-13 & 1600 & $300 d_i\times 194 d_i$  &   $24d_i$& 1.2  & 2812 & 13\% &$45\%$ & N/A\\  
2D-14 & 1600 & $600 d_i\times 388 d_i$  &   $24d_i$& 1.1  & 7913 & 21\% &$53\%$ & N/A\\
2D-15 & 1600 & $1200 d_i \times 776 d_i$ &   $24d_i$& 1.05 & 11220 & 30\% &$66\%$ & N/A\\

 \hline
 \end{tabular*}
 \caption{List of simulation runs with $\sigma \geqslant 6$. 
 The spectral index $p$, the maximum energy ($100$-particle level) at the end of the simulation $\gamma_{max}$, the percentage of magnetic energy that is converted into kinetic energy $E_{kin}\%$, the conversion of magnetic energy caused by perpendicular electric field $(J \cdot E)_\perp$ and $\alpha \tau_{inj}$ estimated by tracking particles in the system.}
 \label{table1}
 \end{center}
\end{table*}

\section{A Simple Model}

It is often argued that some loss mechanism is needed to 
form a power-law distribution 
\citep{Zenitani2001,Drake2010,Drake2013,Hoshino2012}. 
However, the simulation results reported in this paper 
clearly show power-law distributions in a closed periodic 
system. Here we present a simple model to explain  the 
power-law energy spectrum observed in our PIC simulations. 
The model is illustrated by Figure 11(a). As reconnection 
proceeds, cold plasma in the upstream region advects into 
the acceleration zone at a constant velocity that is determined 
by reconnection electric field 
$V_{in} = c\textbf{E}_{rec}\times \textbf{B} /B^2$. The process 
lasts $\tau \sim L_z/2V_{in}$, where $L_z$ is the size of the 
simulation box along the $z$ direction. In the acceleration 
region, our analysis has shown that a first-order Fermi process 
dominates the energy gain during reconnection. We solve the
energy-continuity equation for the energy distribution function 
$f(\varepsilon, t)$ within the acceleration region
\begin{eqnarray}
\frac{\partial f}{\partial t} + \frac{\partial}{\partial \varepsilon} \left( \frac{\partial \varepsilon}{\partial t} f \right) = 0,
\end{eqnarray}

\begin{figure}
\begin{center}
\includegraphics[width=0.5\textwidth]{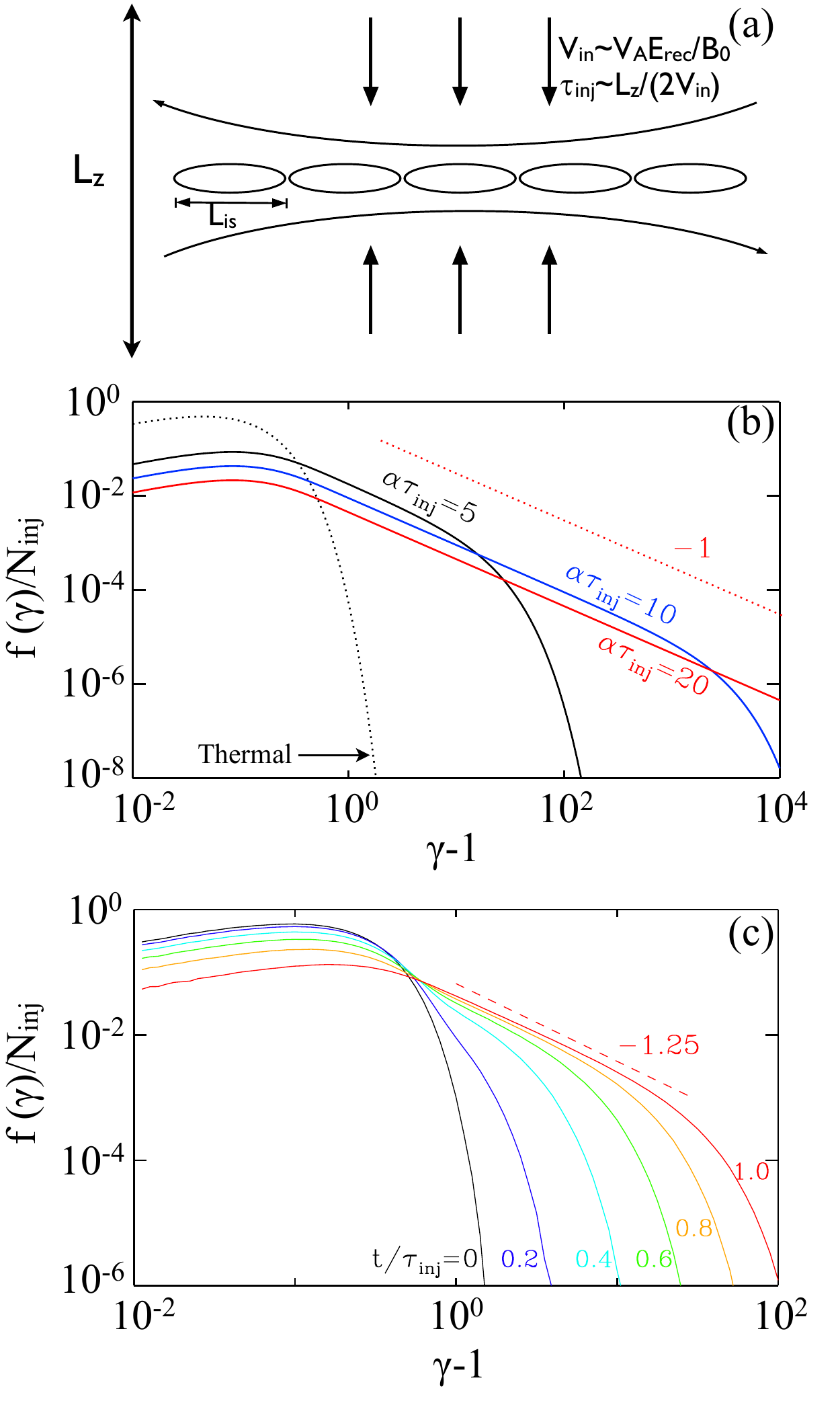}
\caption{(a) Illustration of the acceleration model for the formation of power-law
distributions; (b) Analytical results for different $\alpha \tau_{inj}$ obtained from Eq. (4). (c) The solution of Eq. (1) using time-dependent $\alpha(t)$ from Figure 6 (d).}\label{solution}
\end{center}
\end{figure}

\noindent with acceleration 
$\partial \varepsilon/\partial t = \alpha \varepsilon$, where 
$\varepsilon = m_e c^2 (\gamma-1)/kT$ is the normalized kinetic energy and
$\alpha$ is the constant acceleration rate in the first-order Fermi process. 
We assume that the initial distribution within
the layer $f_0$ is Maxwellian with initial temperature $kT < m_ec^2$, such that
\begin{eqnarray}
f_{0} &\propto & \gamma (\gamma^2-1)^{1/2} \exp (-\varepsilon) \\ \nonumber
        &\approx & \sqrt{2 \varepsilon} \left(1 + \frac{5kT}{4 m_e c^2} \varepsilon + .... \right) \exp(-\varepsilon) \;.
\end{eqnarray}

\noindent For simplicity, we consider the lowest order 
(nonrelativistic) term in this expansion and normalize 
$f_0 = \frac{2N_0}{\sqrt{\pi}}\sqrt{\varepsilon}\exp(-\varepsilon)$ 
by the number of particles
$N_0$ within the initial layer. The distribution after time $t$ is:
\begin{eqnarray} \label{time-solution}
f(\varepsilon, t) = \frac{2N_0}{\sqrt{\pi}} \sqrt{\varepsilon} e^{-3\alpha t/2} \exp(-\varepsilon e^{-\alpha t}),
\end{eqnarray}

\noindent which remains a thermal distribution with a temperature $e^{\alpha t}T$,
consistent with that obtained by \citet{Drake2010}. However, since upstream 
particles enter continuously into the acceleration region, the number of particles 
in the acceleration zone increases with time. 
We consider a particle distribution 
$f_{inj} = \frac{2N_{inj}}{\sqrt{\pi}}\sqrt{\varepsilon}\exp(-\varepsilon)$ 
with number of particles 
$N_{inj} \propto V_{in}\tau_{inj}$ injected from upstream, 
where $\tau_{inj}$ is the time scale for particle injection.
To highlight the key role that time-dependent injection plays in setting up the power-law, we first consider a quick heuristic derivation of the main result. 
To proceed, we split $f_{inj}$ into $N$ groups and release the $j^{th}$ group into the acceleration region at time $t = j \Delta t$.
Since each group will satisfy the Equation \ref{time-solution} for a different initial time, 
after we have injected the final group at $t = \tau_{inj}$, the total distribution (in the limit
$N \to \infty$) is
\begin{eqnarray}\label{integral}
f(\varepsilon, t) &\sim& \frac{2N_{inj}}{\sqrt{\pi}\tau_{inj}} \int^{\tau_{inj}}_0 
\sqrt{\varepsilon}e^{-3\alpha t/2} \exp{(-\varepsilon e^{-\alpha t})} dt \\ \nonumber
 &=& \frac{N_{inj}}{\alpha \tau_{inj}} [ 
 \frac{\erf(\varepsilon^{1/2}) - \erf(\varepsilon^{1/2}e^{-\alpha \tau_{inj}/2})}{\varepsilon}  \\ 
 &+& \frac{2}{\sqrt{\pi}} \frac{e^{-\alpha \tau_{inj} /2}\exp(-\varepsilon e^{-\alpha \tau_{inj}}) - e^{-\varepsilon}}{\varepsilon^{1/2}}] \nonumber
\end{eqnarray}

 
\noindent In the limit of $\alpha \tau \gg 1$, this gives the relation 
$f \propto 1/\varepsilon$ in the energy range $1 < \varepsilon < e^{\alpha \tau_{inj}}$. Figure \ref{solution}(b) shows (4) for different $\alpha \tau_{inj}$.
A power-law spectrum with $p = 1$ emerges as $\alpha \tau_{inj}$ increases $\alpha \tau_{inj} > 1$. 
Note that for a 
closed system, since the averaged magnetic energy per 
particle is only $\sigma m_e c^2/4$ and the energy in each energy
bin is constant, the maximum energy 
the power law can only extend to $\gamma_{max} \sim \sigma/4$.
 
Next, in order to treat the problem more rigorously, and include the influence 
of particle escape, we consider the more complete equation
\begin{eqnarray}
\frac{\partial f}{\partial t} + \frac{\partial}{\partial \varepsilon} \left( \frac{\partial \varepsilon}{\partial t} f \right) = \frac{f_{inj}}{\tau_{inj}} - \frac{f}{\tau_{esc}},
\end{eqnarray}

\noindent where $\tau_{esc}$ is the escape time for particles. For the initial current-layer distribution $f_0$ and injected particle 
distribution $f_{inj}$ considered above,
the solution can be written as 
\begin{eqnarray} \label{complete-solution}
 f(\varepsilon,t) &=& \frac{2 N_0}{\sqrt{\pi}} \sqrt{\varepsilon} e^{-(3/2+\beta ) \alpha t} \exp(-\varepsilon e^{-\alpha t}) \\ \nonumber
 &+& \frac{2 N_{inj}}{\sqrt{\pi}(\alpha \tau_{inj}) \varepsilon^{1+\beta}} \left[ 
 \Gamma_{(3/2+\beta)}(\varepsilon e^{-\alpha t}) - \Gamma_{(3/2+\beta)}(\varepsilon) \right] ,
\end{eqnarray} 

\noindent where $\beta = 1/(\alpha \tau_{esc})$ and $\Gamma_s (x)$ is the 
incomplete Gamma function. The first term accounts for particles initially in the acceleration region while the 
second term describes the evolution of injected particles. In the limit of no injection or 
escape ($\tau_{esc} \to \infty$ and $\tau_{inj} \to \infty$), the first term in 
(\ref{complete-solution}) remains a thermal distribution the same as (3). 
However, as reconnection proceeds new particles 
enter continuously into the acceleration region and due to the periodic boundary conditions there is no particle escape. Thus considering the case $\tau_{esc} \to \infty$ and assuming $N_0 \ll N_{inj}$, at the time $t=\tau_{inj}$ when reconnection saturates the second term in (6) simplifies to (4).
Thus in the limit $N_0 \sim N_{inj}$ the first term in (6) should be 
retained and the power-law produced is sub-thermal relative to this population. While it is 
straightforward to obtain the relativistic corrections arising from the injected 
distribution (2), we emphasize that these terms do not alter the spectral 
index. This solution explains results from our simulations, and also appears to
explain the results from several recent papers, which obtained power-law
distributions by subtracting the initial hot plasma component in the current layer 
\citep{Sironi2014,Melzani2014b,Werner2014}.
In particular, \citet{Melzani2014b} explicitly discussed the evolution of particle
distribution initially in the current layer and reported it as a heated 
Maxwellian distribution.

In order to estimate the acceleration rate $\alpha$, the energy change of each particle can be approximated by a relativistic collision formula \citep[e.g.,][]{Longair1994}
\begin{eqnarray}
\Delta \varepsilon =\left(\Gamma_V^2(1+\frac{2Vv_x}{c^2}+\frac{V^2}{c^2}) - 1 \right)\varepsilon,
\end{eqnarray}

\noindent where $V$ is the outflow speed, $\Gamma_V^2 = 1/(1-V^2/c^2)$, and 
$v_x$ is the particle velocity in the $x$ direction. The time between two 
collisions is about $L_{is}/v_x$, where $L_{is}$ is the typical size of the 
magnetic islands (or flux ropes in 3D). Assuming that relativistic particles 
have a nearly isotropic distribution $v_x \sim c/2$, then
\begin{eqnarray}
\alpha = \frac{\Delta \varepsilon}{\varepsilon \Delta t} \sim \frac{c(\Gamma_V^2(1+\frac{V}{c}+\frac{V^2}{c^2})-1)}{2L_{is}}.
\end{eqnarray}
\noindent Using this expression, we measure the averaged $V$ and $L_{is}$ from the
simulations and estimate the time-dependent acceleration rate $\alpha (t)$. An example is shown in Figure 6 (d). This agrees reasonably well with that obtained from perpendicular acceleration and curvature drift acceleration. 
Figure 7(c) shows the time-integrated value of $\alpha \tau_{inj} = \int^{\tau_{inj}}_0 \alpha(t) dt$ for various simulations with $\sigma = 6 - 400$.  For cases with $\alpha \tau_{inj} > 1$, a hard power-law distribution with spectral index $p \sim 1$ forms. 
For higher $\sigma$ and larger system size, the magnitude of $\alpha \tau_{inj}$ 
increases approximately as $\propto \sigma^{1/2}$. 
 
Better agreement between the simple model and the PIC
simulations can be reached by considering the 
time-dependent acceleration rate $\alpha(t)$.
As the magnetic reconnection saturates, the acceleration rate decreases. Figure 11 (c) shows
the solution that uses the time-dependent acceleration
rate $\alpha(t)$ in Figure 6(d) using 
a stochastic integration technique described by  
\citet{Guo2010}. 
The final spectral index is 
about $p = 1.25$,
similar to that from the PIC simulation
shown in Figure 7(a).

\section{Implications}

We discuss the implication of the above conclusions
for understanding the role of magnetic reconnection
in magnetically dominated astrophysical
systems. Based on the current understanding of magnetic 
reconnection, multiple X-line reconnection develops when
the secondary tearing instability is active in large-scale 
collisionless plasma system. This process may also be important 
when a hierarchy of collisional plasmoids \citep{Loureiro2007,Bhattacharjee2009,Uzdensky2010} 
develops kinetic
scale current layers that may trigger collisionless reconnection \citep{Daughton2009,Ji2011}. Therefore the collisionless 
reconnection process discussed here is relevant to 
a range of high-energy astrophysical problems below \citep[see][for a comprehensive summary of astrophysical problems with relevant physics]{Ji2011}.

\subsection{Pulsar Wind Nebulae}

In PWN models, magnetic reconnection has been proposed as a 
mechanism for dissipating magnetic energy in Poynting-flux
dominated flows
\citep{Coroniti1990,Lyubarsky2001,Kirk2003,Porth2013} and 
accelerating particles to high energies \citep{Kirk2004}. In 
PWNe, the emission flux usually has spectral indices 
$\alpha_\nu = 0 - 0.3$ in the radio range, which requires an 
electron energy distribution $dN/d\gamma \propto \gamma^{-p}$ 
with $p = 1 - 1.6$ ($p = 2 \alpha_\nu + 1$), too hard to be 
explained by diffusive shock acceleration \citep{Atoyan1999}. 
The recently detected $>100$-MeV Crab flares have photon 
energies well above the usually employed upper limit for 
synchrotron emissions, challenging the traditional 
acceleration theory \citep{Abdo2011,Tavani2011,Buhler2014}. 
There are two main possibilities for explaining the photon 
energies, (1) a relativistic Doppler boosting of the emitting 
region \citep{Clausen2012} and/or (2) a strong particle 
acceleration in a nonideal electric field where $E > B_\perp$, 
where $B_\perp$ is the magnetic field perpendicular to particle 
velocity \citep{Cerutti2013,Lyutikov2014}.

These observations suggest that relativistic 
magnetic reconnection may occur in the Crab nebula.  The power law
index revealed in this study is $p = 1 - 2$, consistent with 
the inferred spectra in the radio range \citep{Atoyan1999} and 
in high-energy during the Crab $\gamma-$ray flares 
\citep{Tavani2011}. Explaining these observations requires a 
fast and efficient dissipation mechanism that converts a 
substantial fraction of magnetic energy into relativistic particles
\citep{Lyutikov2014}. In the Crab pulsar, magnetic reconnection 
is estimated to be in the plasmoid dominated 
regime and can dissipate a nontrivial fraction of the pulsar spin-down power
\citep{Uzdensky2014}. Our simulations have shown that for a magnetically dominated 
reconnection layer with $\sigma \gg 1$, magnetic reconnection rate is
greatly enhanced by about one order of magnitude compared to the nonrelativistic
limit \citep[see also][]{Liu2015} and a large fraction of magnetic energy in the system is converted into nonthermal energy distribution, suggesting an efficient magnetic dissipation and strong nonthermal
radiation processes
in the Crab wind. The maximum particle energy increases linearly and can be well predicted 
by assuming particles moving along the reconnecting electric field at 
the speed of light.
There are also relativistic inflow and outflow structures ($\Gamma_{max} \gtrsim 10$) 
associated with 
reconnection, which may boost the emission photon energy and help to explain the observed Crab flares \citep{Clausen2012}.
It is interesting to note that the reconnection acceleration
may also explain the pulsed $\gamma$-ray emission, although 
observations at higher energies is required to further constrain the model \citep{Mochol2015}.

\subsection{AGN Jets}

In AGN jets, a number of $\gamma$-ray sources have flat radio spectra with indices
around $\alpha_\nu = 0$, meaning the electron energy distribution index may be close to $p=1$ \citep{Abdo2010,Hayashida2015}.
Several blazars have shown extremely fast variability in TeV range 
  on the order of several minutes 
\citep{Aharonian2007,Albert2007}. 
Hard power laws $p \sim 1$ in TeV range have been
inferred after removing effect of the extragalactic background light using various models \citep{Aharonian2006,Krennrich2008}. For GeV-TeV flat spectrum radio quasars (FSRQ), high radiation efficiency is reported \citep{Zhang2013} and the electron $\sigma_e$
which is measured as magnetic energy power to the electron energy power is very high up to the order of $100$ \citep{Zhang2014}.

Explaining the fast variability requires 
the relativistic beaming effect possibly arising from relativistic 
reconnection outflows
\citep{Giannios2009,Deng2014}. Our kinetic simulations have shown 
that the Lorentz factor of the maximum outflow speed 
$\Gamma_x \sim 10$ for $\sigma \sim 1000$. The simulation 
results and theoretical model predict hard particle 
energy distribution consistent with
the hard radio spectra observed in some AGNs \citep{Romanova1992}.
Recent advanced AGN emission models have inferred that at least 
for some types of blazers, particularly 
FSRQ, strong particle acceleration and/or strong magnetic
field is necessary to explain fast flares and $\sigma$ inferred from the model fitting can significantly 
exceed unity $\sigma \gg 1$ \citep{Chen2014}.
Magnetic reconnection may offer an explanation for the simultaneous decrease of magnetic field 
and emission increase during the flare phase of blazar flares
and is a promising scenario for modeling AGN emissions \citep{Zhang2014}.

\subsection{Gamma-ray bursts}

In gamma-ray bursts (GRBs), the 
traditional internal shock model of prompt
emission is difficult
to reconcile with observations \citep[see][and references therein]{Zhang2011}.
Magnetic reconnection and associated particle acceleration have been proposed as a key 
process in GRB models such as ICMART model \citep{Zhang2011} and reconnection-switch model 
\citep{McKinney2012}. 
The efficient magnetic dissipation and particle acceleration 
during reconnection may be important to 
understand the emission mechanism in GRBs \citep{Kumar1999,Spruit2001,Drenkhahn2002}. 
\citet{Gruber2014} have shown a series of features in GRB prompt 
emission that are not consistent with the simple synchrotron shock 
model. For example, the hard low-energy spectra, where the 
particle energy spectral index is close to $p = 1$ assuming synchrotron radiation \citep{Ghisellini2000,Preece2002} and the
thermal emission component predicted in
the fireball-internal-shock model has been rarely seen in GRBs
\citep{Zhang2009,Abdo2009}.

From our simulation results and analytical model, the particle 
energy spectral index
is close to $p = 1$, consistent with low-energy photon spectra 
observed in most GRBs \citep{Band1993,Preece2000,Gruber2014}.
The acceleration in reconnection layers is much faster than
the radiation cooling and can maintain the hard spectrum.
Using PIC simulation, \citet{Spitkovsky2008} found that in the downstream region of highly relativistic shocks the number of particles in the nonthermal tail is $\sim 1\%$ of the entire downstream population, and they carry $∼10\%$ of the kinetic energy in the downstream
region. In our simulations of relativistic reconnection, the number of 
nonthermal relativistic particles is $\sim 25\%$ of the total number 
particles in the simulation and they carry $\sim 95\%$ of kinetic energy 
in the system, meaning relativistic reconnection is much more efficient in producing nonthermal relativistic particles. This efficient conversion from magnetic energy into kinetic energy of nonthermal particles may help solve the efficiency problem in GRB models \citep{Zhang2007,Deng2014}.

\subsection{Nonrelativistic reconnection sites}

While the primary focus of this paper is relativistic magnetic reconnection, the physics of Fermi acceleration and the formation of power-law distribution is also applicable to the nonrelativistic 
regimes previously discussed 
\citep{Drake2006,Drake2010,Drake2013}. Based on our analytical 
model, the power-law distribution forms only when $\alpha \tau_{inj} > 1$. 
The results in this paper demonstrate that this condition 
is more easily achieved in regimes with $\sigma \gg 1$, but 
it may also occur with $\sigma < 1$ in sufficiently 
large reconnection layers. In several preliminary simulations, 
we have observed the formation of similar
power laws in nonrelativistic proton-electron plasma and
will report elsewhere. 

X-ray observations of solar flares have shown strong 
particle acceleration and energy conversion during magnetic reconnection and the particle 
distribution often takes power-law distributions, requiring 
a particle acceleration mechanism that is dominated by nonthermal
acceleration
\citep{Krucker2008,Krucker2010,Krucker2014}. 
As we have shown here, in magnetically dominated regimes, a large fraction of magnetic energy can
be converted into particles in a power-law distribution. Similar process
is likely to occur in solar flares, 
where the plasma $\beta = 8\pi nkT/B^2 \sim 0.001$ - $0.01$ ($\sigma < 1$).
However, physics such as the influence of $m_i/m_e$, strong trapping at X-line region, and particle escape from the system
need to be investigated further (Egedal \& Daughton 2015, in preparation).

\section{Discussion and Conclusion}

The dissipation of magnetic field and particle energization 
in the magnetically dominated systems is of strong interest in 
high energy astrophysics. In this study, we use 2D and 3D 
fully kinetic simulations that resolve the full range of 
plasma physics to 
investigate the particle acceleration and 
plasma dynamics during collisionless magnetic reconnection in 
a pair plasma with magnetization parameter $\sigma$ 
varying from $0.25$ to $1600$. A force-free current 
layer, which does not require a hot plasma population 
in the current layer, is implemented as the initial 
condition.

We find that the evolution of the current sheet and 
acceleration of particles has two
stages. In the early stage, an extended reconnection 
region forms and generates 
a parallel electric field that accelerates particles in the current layer. 
As time proceeds, the layer breaks into multiple plasmoids 
(flux ropes in 3D) due to the secondary tearing instability. 
The motional electric
field in the reconnection layer strongly accelerates 
energetic particles via a first-order relativistic 
Fermi process leading to the conversion of 
most of the free energy in the system.
A large fraction of the magnetic
energy is quickly converted into the kinetic energy of nonthermal relativistic 
particles (within a few light-crossing times) and the eventual energy spectra show a power law $f \propto (\gamma-1)^{-p}$,
with the spectral index $p$ decreasing with $\sigma$ and system size and approaching $p = 1$.
The formation of the power-law distribution can be described by a simple model
that includes both inflow and the Fermi acceleration. This model also 
appears to explain recent PIC simulations \citep{Sironi2014,Melzani2014b,Werner2014}, 
which reported hard
power-law distributions after subtracting the initial hot plasma population inside
the current layer. 
For the more realistic limit with both particle loss and injection, the spectral
index $p = 1 + 1/(\alpha \tau_{esc})$, recovering the classical Fermi solution.
If the escape is caused by convection out of the reconnection region 
$\tau_{esc} = L_x /V_x$, the spectral index should approach $p = 1$ when
$\alpha \tau_{esc} \gg 1$ in the high-$\sigma$ regime. In preliminary 2D 
simulations using open boundary conditions, we have confirmed this trend
and will report elsewhere. For the nonrelativistic limit, the reconnection
needs to be sustained over a longer time to form a power law.

We have also shown that in sufficiently high-$\sigma$ regimes 
the magnetic reconnection rate is enhanced and relativistic 
inflow and outflow structures develop. The scaling follows the prediction 
based on the 
Lorentz contraction of plasma passing through the diffusion region. 
Although 3D magnetic turbulence is generated as a consequence 
of the growth of the secondary tearing instability and kink instability, 
the particle acceleration, energy release and reconnection 
rate in the 3D simulation are comparable to the corresponding
2D simulation. 

Our study has demonstrated that relativistic magnetic 
reconnection is a highly efficient energy-dissipation 
mechanism in the magnetically dominated regimes.
The plasma distribution in the reconnection layer features 
power-law energy spectra, which may be important in 
understanding the nonthermal emissions from objects like 
pulsars, jets from black holes, and gamma-ray bursts.
Both the inflow and outflow speeds approach the speed 
of light and have Lorentz factors of a few, which
may explain the fast variability and high luminosity 
observed in those high-energy astrophysical systems. 
These findings on particle acceleration and plasma dynamics 
during relativistic reconnection substantiate the important role 
of magnetic reconnection in high-energy astrophysical systems.


\section*{Acknowledgement}
We gratefully acknowledge useful 
discussions with and comments from Andrey Beresnyak, 
Xuhui Chen, Wei Cui, Wei Deng, Brenda Dingus,  Jim Drake, 
Joe Giacalone, Dimitrios Giannios, Serguei Komissarov, Pawan Kumar,
Xiaocan Li, Maxim Lyutikov, Rob Preece, Marc Swisdak, 
Alexander Tchekhovskoy, Dmitri Uzdensky, Yajie Yuan, 
Gary Zank, Bing Zhang, and Haocheng Zhang.
This work is supported by the DOE through the LDRD program 
at LANL and DOE/OFES support to LANL in collaboration with 
CMSO. The research is part of the Blue Waters sustained-petascale 
computing project, which is supported by the NSF (Grand No. 
OCI 07-25070) and the state of Illinois. Additional simulations 
were performed at the National Center for Computational Sciences 
at ORNL and with LANL institutional computing.

\section*{Appendix: Numerical Convergence}
The accuracy of particle-in-cell (PIC) kinetic 
simulations depends on
a series of numerical parameters such as cell size, 
time step, and the number of
macro-particles per cell \citep[e.g.,][]{Birdsall1991}. 
The numerical convergence of simulation results has
been rarely explicitly 
checked when modeling
astrophysical problems using PIC simulations, 
and often a small 
number of macro-particles are used.
Here we examine the numerical convergence of our results 
on these numerical parameters using VPIC code for different initial 
temperatures from $kT_0 = 0.01$ to $0.36$
$m_e c^2$ . Our test case has $\sigma = 25$ with box size 
$L_x \times L_z = 600 d_i \times 388 d_i$ and simulation 
time $\omega_{pe}t = 3000 $.
We find that numerical heating can become unacceptably large 
when a small number of particles per cell is used.
In Table 2 we list the key parameters for the test.
Although for most cases, the violation in energy conservation 
is small ($E_{err}/E_{tot}$ within $1\%$), the numerical heating
can significantly modify the particle distribution
since the initial kinetic energy is a small fraction of the total energy. Therefore
to obtain trustworthy results that are numerically converged, 
the violation of energy conservation should be much less than
the initial kinetic energy $E_{err}/E_{k0} \ll 1$. 
Figure 13 shows several cases with $kT_0 = 0.36 m_e c^2$ 
with grid number $2048 \times 2048$, Courant number $C_r = 0.7$, and different 
numbers of particles per cell from $8$ to $512$.
Figure 14 shows several cases for $kT_0 = 0.01 m_e c^2$ 
with grid number $4096 \times 4096$ and $C_r = 0.9$ but 
different numbers of particles per cell from 2 to 512. Both 
figures show that as the total energy change in the 
numerical simulations becomes smaller than the initial
kinetic energy $E_{err}/E_{k0} \ll 1$, the numerical heating has a negligible 
effect on the distribution function.

 \begin{figure}
 \begin{center}
 \includegraphics[width=0.45\textwidth]{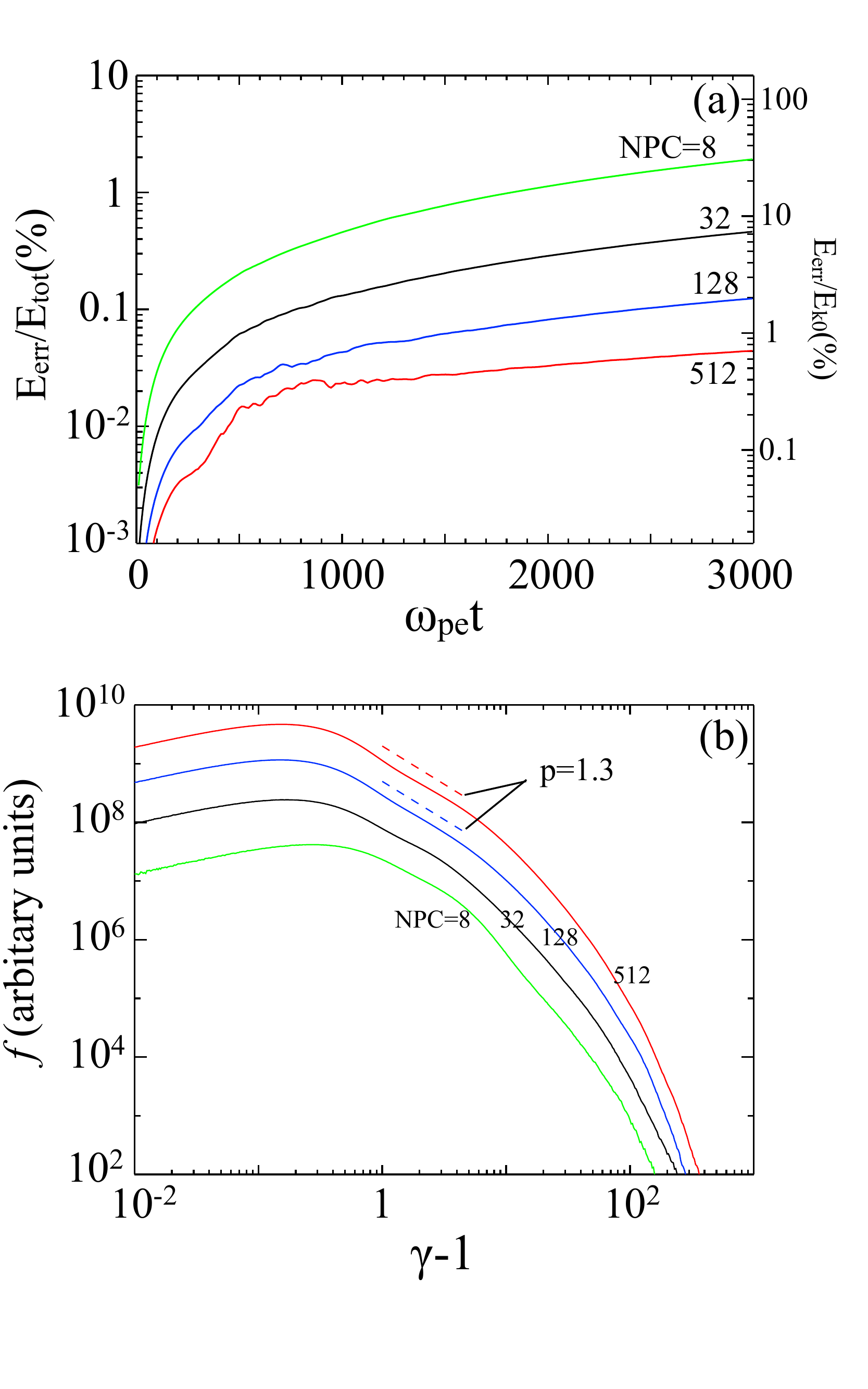}
\caption{Several cases with $kT_0 = 0.36 m_e c^2$ with grid number $2048 \times 2048$, Courant number $C_r = 0.7$ but different numbers of particles per cell from 8 to 512. \label{Fig-energyconservation}}
 \end{center}
 \end{figure}
 
  \begin{figure}
 \begin{center}
 \includegraphics[width=0.45\textwidth]{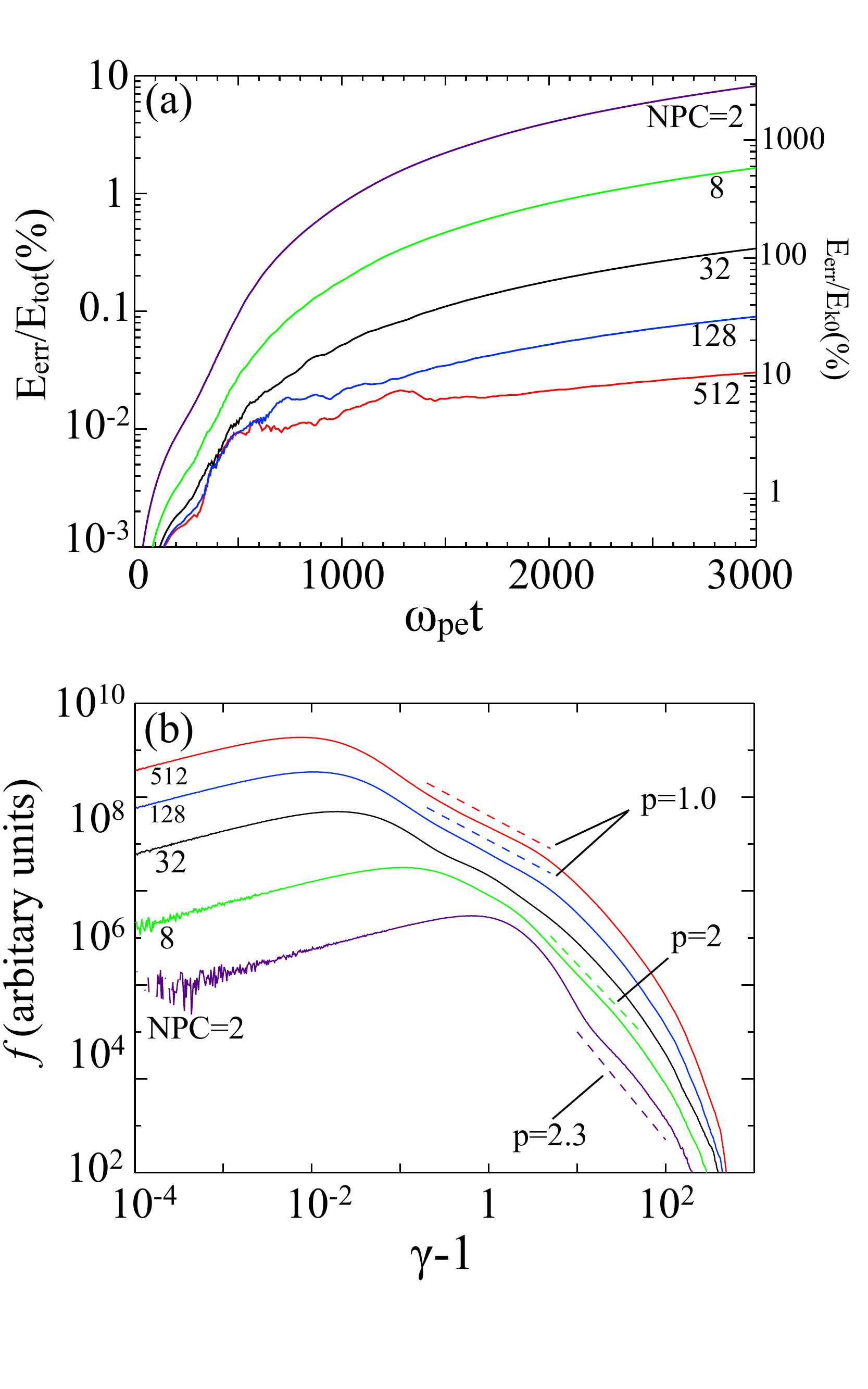}
\caption{Several cases for $kT_0 = 0.01 m_e c^2$ 
with grid number $4096 \times 4096$ and $C_r = 0.9$ but 
different numbers of particles per cell from 2 to 512.\label{Fig-energyconservation}}
 \end{center}
 \end{figure}
 

\begin{table*}
\begin{center}
\begin{tabular}{|c|cccc|cc|}
\hline 
\textbf{Run} & $kT_0/m_ec^2$ & Grid numbers & Time step (Cr) &  NPC & $E_{err}/E_{total}$ & $E_{err}/E_{k0}$  \\
\hline
A-1 & $0.36$  & $4096 \times 4096$    & 0.9 &   2 &  $10\%$  & $159 \%$ \\

A-2 & $0.36$  & $4096 \times 4096$    & 0.9 &   8 &  $2.4\%$  & $38 \%$ \\

A-3 & $0.36$  & $4096 \times 4096$    & 0.9 &   32 &  $0.56\%$  & $9 \%$ \\

A-4 & $0.36$  & $4096 \times 4096$    & 0.9 &   128 &  $0.084\%$  & $1.3 \%$ \\

A-5 & $0.36$  & $2048 \times 2048$    & 0.9 &   8 &  $5\%$  & $80\%$ \\

A-6 & $0.36$  & $2048 \times 2048$    & 0.9 &   32 &  $1.2\%$  & $20\%$ \\

A-7 & $0.36$  & $2048 \times 2048$    & 0.9 &   128 &  $0.3\%$  & $5\%$ \\

A-8 & $0.36$  & $2048 \times 2048$    & 0.7 &   8 &  $1.9\%$  & $30\%$ \\

A-9 & $0.36$  & $2048 \times 2048$    & 0.7 &   32 &  $0.45\%$  & $7\%$ \\

A-10 & $0.36$  & $2048 \times 2048$    & 0.7 &   128 &  $0.12\%$  & $1.9\%$ \\

A-11 & $0.36$  & $2048 \times 2048$    & 0.7 &   512 &  $0.04\%$  & $0.6\%$ \\

A-12 & $0.36$  & $2048 \times 2048$    & 0.5 &   8 &  $0.75\%$  & $12\%$ \\

A-13 & $0.36$  & $2048 \times 2048$    & 0.5 &   32 &  $0.19\%$  & $3\%$ \\

A-14 & $0.36$  & $2048 \times 2048$   & 0.5 &   128 &  $0.05\%$  & $0.8\%$ \\

B-1 & $0.09$ & $4096 \times 4096$    & 0.9 &   2 &  $9\%$  & $474 \%$ \\

B-2 & $0.09$  & $4096 \times 4096$    & 0.9 &   8 &  $1.9\%$  & $100 \%$ \\

B-3 & $0.09$  & $4096 \times 4096$    & 0.9 &   32 &  $0.42\%$  & $22 \%$ \\

B-4 & $0.09$  & $4096 \times 4096$    & 0.9 &   128 &  $0.05\%$  & $2.6 \%$ \\

B-5 & $0.09$  & $2048 \times 2048$    & 0.9 &   8 &  $4.2\%$  & $212 \%$ \\

B-6 & $0.09$  & $2048 \times 2048$    & 0.9 &   32 &  $0.9\%$  & $45 \%$ \\

B-7 & $0.09$  & $2048 \times 2048$    & 0.9 &   128 &  $0.24\%$  & $12 \%$ \\

B-8 & $0.09$  & $2048 \times 2048$    & 0.7 &   8 &  $1.6\%$  & $80 \%$ \\

B-9 & $0.09$  & $2048 \times 2048$    & 0.7 &   32 &  $0.37\%$  & $19 \%$ \\

B-10 & $0.09$  & $2048 \times 2048$    & 0.7 &   128 &  $0.1\%$  & $5 \%$ \\

B-11 & $0.09$  & $2048 \times 2048$    & 0.5 &   8 &  $0.6\%$  & $30 \%$ \\

B-12& $0.09$  & $2048 \times 2048$    & 0.5 &   32 &  $0.13\%$  & $7 \%$ \\

B-13 & $0.09$  & $2048 \times 2048$    & 0.5 &   128 &  $0.04\%$  & $2 \%$ \\

C-1 & $0.01$  & $4096 \times 4096$  & 0.9 & 2 &  $8.5\%$  & $3000 \%$ \\

C-2 & $0.01$  & $4096 \times 4096$  & 0.9 & 8 &  $1.7\%$  & $595 \%$  \\

C-3 & $0.01$  & $4096 \times 4096$    & 0.9 &   32 &  $0.36\%$  & $126 \%$ \\

C-4 & $0.01$  & $4096 \times 4096$    & 0.9 &   128 &  $0.09\%$  & $32 \%$ \\

C-5 & $0.01$  & $4096 \times 4096$    & 0.9 &   512 &  $0.03\%$  & $10 \%$ \\

C-6 & $0.01$  & $2048 \times 2048$    & 0.9 &   32 &  $0.75\%$  & $265\%$ \\

C-7 & $0.01$  & $2048 \times 2048$    & 0.9 &   128 &  $0.19\%$  & $67\%$ \\

C-8 & $0.01$  & $2048 \times 2048$    & 0.9 &   512 &  $0.08\%$  & $29\%$ \\

C-9 & $0.01$  & $2048 \times 2048$    & 0.7 &   32 &  $0.28\%$  & $102\%$ \\

C-10 & $0.01$  & $2048 \times 2048$    & 0.7 &   128 &  $0.08\%$  & $28\%$ \\

C-11 & $0.01$  & $2048 \times 2048$    & 0.7 &   512 &  $0.044\%$  & $15\%$ \\

C-12 & $0.01$  & $2048 \times 2048$    & 0.5 &   32 &  $0.11\%$  & $39\%$ \\

C-13 & $0.01$  & $2048 \times 2048$    & 0.5 &   128 &  $0.035\%$  & $12\%$ \\

C-14 & $0.01$  & $2048 \times 2048$    & 0.5 &   512 &  $0.014\%$  & $5\%$ \\

 \hline
 \end{tabular}
 \caption{List of simulation runs used to test numerical convergence. All the runs
are for $\sigma = 25$ and $L_x \times L_z = 600 d_i \times 388 d_i$ and were performed
over a duration $\omega_{pe}t = 3000$. Note $kT_0/m_ec^2$ 
is the initial plasma temperature normalized by rest energy $m_e c^2$. Time step is 
represented by the dimensionless Courant number $C_r = c\Delta t / \Delta r$, where 
$\Delta r = \Delta x \Delta y \Delta z / (\Delta x \Delta y + \Delta y \Delta z + \Delta x \Delta z)$. 
NPC represents the number of particle pairs per cell. $E_{err}/E_{total}$ represents the 
ratio between change of total energy compare to the initial total energy. 
$E_{err}/E_{k0}$ represents the ratio between change of total energy compare to the 
initial plasma kinetic energy.}
 \label{table2}
 \end{center}
\end{table*}


\end{document}